\documentclass{article}

\usepackage{ijcai22}

\usepackage{times}

\usepackage{soul}
\usepackage{url}
\usepackage[utf8]{inputenc}
\usepackage[small]{caption}
\usepackage{graphicx}
\usepackage{amsmath}
\usepackage{booktabs}
\usepackage{xspace}

\usepackage{comment}

\usepackage{color}
\usepackage{pstricks,pst-node,pst-tree,pstricks-add}
\usepackage{multido}
\usepackage{authblk}

\usepackage{enumerate}
\usepackage[english]{babel}
\usepackage{amsthm}
\usepackage{amssymb}
\usepackage{amsmath}
\usepackage{amsfonts}
\usepackage[footnote,marginclue]{fixme}
\usepackage{appendix}
\usepackage{enumitem}

\usepackage{algorithm}
\usepackage[noend]{algpseudocode}
\newcommand{\return}{{\mathbf{output}}}
\newcommand{\accept}{{\mathsf{true}}}
\newcommand{\reject}{{\mathsf{false}}}
\newcommand{\false}{{\mathsf{false}}}

\newcommand{\guess}{{\mathbf{Guess}}}

\theoremstyle{plain}
\newtheorem{theorem}{Theorem}
\newtheorem{lemma}[theorem]{Lemma}
\newtheorem{corollary}[theorem]{Corollary}
\newtheorem{proposition}[theorem]{Proposition}

\theoremstyle{definition}

\newtheorem{definition}[theorem]{Definition}
\newtheorem{example}[theorem]{Example}

\newcommand{\sg}{\sigma}

\newcommand{\mA}{{\mathfrak A}}
\newcommand{\Dom}{{\rm Dom}}
\newcommand{\dom}{{\rm dom}}

\newcommand{\restrict}[2]{{#1}({#2})}

\newcommand{\FO}[1]{{\rm FO(#1)}}

\newcommand{\free}[1]{\textsf{free}(#1)}
\newcommand{\subformula}[1]{\textrm{SubF}(#1)}
\newcommand{\fo}{{\rm FO}}

\newcommand{\Fr}{{\rm Fr}}
\newcommand{\Var}{{\rm Var}}

\newcommand{\ESO}{{\rm ESO}}

\newcommand*\dep{{=\mkern-1.2mu}} 

\newcommand{\FOdep}{{\rm FO}(\dep())}
\newcommand{\df}{\fo(\mathcal{D})}

\newcommand{\FOinclusion}{{\rm FO(\subseteq)}}
\newcommand{\FOanon}{\fo({\rm \Upsilon})}
\newcommand{\anon}[2]{{#1}\hspace{1pt}{\Upsilon}\hspace{1pt}{#2}}
\newcommand{\const}{\dep{(\cdot)}}

\newcommand{\np}{{\rm NP}}
\newcommand{\Ptime}{{\rm PTIME}}

\newcommand{\nl}{{\rm NL}}
\newcommand{\logspace}{{\rm Logspace}}

\newcommand{\PGFP}{{\rm GFP}^+}

\newcommand{\tu}[1]{\overline{#1}}

\newcommand{\indep}[3]{{#1}\ \bot_{#2}\ {#3}}
\newcommand{\var}[1]{\mathsf{Var(#1)}}

\usepackage{mathrsfs}
\newcommand{\calL}{\mathcal{L}}

\newcommand{\calS}{\mathcal{S}}
\newcommand{\calC}{{\mathcal C}}

\newcommand{\scrM}{\mathscr{M}}

\newcommand{\scrS}{\mathscr{S}}

\newcommand{\mX}{{\mathfrak X}}

\newcommand{\sfX}{\mathsf{X}}

\newcommand{\commenttext}[1]{ \begin{center} {\fbox{\begin{minipage}[h]{0.9
            \linewidth}   {\sf #1} \end{minipage} }} \end{center}} 

\newcommand{\arnaud}[1]{ {\leavevmode\color{blue}\commenttext{Arnaud: #1}}}
\newcommand{\juha}[1]{{\color{blue}\commenttext{Juha: #1}}}
\newcommand{\jouko}[1]{{\color{blue}\commenttext{Jouko: #1}}}

\newcommand{\newaddition}[1]{{\color{orange}\commenttext{New addition: #1}}}

\newcommand{\longversion}[1]{#1}
\newcommand{\shortversion}[1]{}

 \newcommand{\pb}[1]{\textsc{#1}}

\newcommand{\Disjunctiondepth}[1]{\textsf{d}_{\vee}(#1)}

\newcommand{\indlogic}{\rm{FO} (\bot_{\rm c})}

\newcommand{\bcindepfo}{\mathsf{BC}(\bot,\fo)}
\newcommand{\CCm}{\mathfrak{C}^{-}}
\newcommand{\CCp}{\mathfrak{C}^{+}}

\title{Modular SAT-based techniques for reasoning tasks  in team semantics}

\begin{document}

	\author[1]{Arnaud Durand}
	\author[2]{Juha Kontinen}
	\author[2]{Jouko V\"{a}\"{a}n\"{a}nen}
	\affil[1]{IMJ-PRG, CNRS UMR 7586 - Universit\'e de Paris, France}
	\affil[2]{Department of Mathematics and Statistics, University of Helsinki, Finland} 

	\maketitle

\begin{abstract}
We study the complexity of reasoning tasks for logics in team semantics. Our main  focus is on the data complexity of model checking but we also derive new results for logically defined counting and enumeration problems. Our approach is based on modular reductions of these problems into the corresponding problems of various  classes of Boolean formulas. We illustrate our approach via several new tractability/intractability results.
\end{abstract}
\section{Introduction}

Designing languages that are both flexible and highly expressive but that also admit fast and efficient algorithmic procedures is an important goal in knowledge representation. Team semantics supports a plethora of new languages in which various data dependences are treated as first-class citizens and are thus well suited for knowledge representation and reasoning. In this article we study the complexity of logically defined reasoning tasks for various logics in team semantics. Our novel approach reduces questions such as decision, counting and enumeration of teams satisfying a formula in a modular fashion into the corresponding problems for various classes of Boolean clauses. 



Our starting point is first order logic but we enhance it by adding new atomic expressions corresponding to various data dependencies. Dependence logic \cite{vaananen07}  extends first-order logic by dependence atoms of the form 
$\dep(x_1,\ldots,x_n)$
expressing that the value of $x_n$ is functionally determined by the values of the variables $x_1,\ldots, x_{n-1}$.  Independence  and inclusion logic  \cite{gradel10,galliani12} are variants of  dependence logic 
that replace dependence atoms by  independence and inclusion atoms 
$\tu{y}\ \bot_{\tu{x}} \ \tu{z}$ and  $\tu{x}\subseteq \tu{y},$
respectively. Dependence, inclusion, and independence  atoms are intimately connected to the corresponding functional, inclusion, and multivalued  dependences in relational databases. Independence atoms can be also viewed as the qualitative analogue of the notion of conditional independence in statistics  \cite{DBLP:journals/networks/GeigerVP90}. In fact, all of the first-order operators in team semantics can be also  naturally given a probabilistic interpretation resulting with a probabilistic version of team semantics in which probabilistic and qualitative dependencies can be studied and applied in a uniform framework \cite{HannulaHKKV19,HKMV18}.

Dependence logic and its variants have been used to formalize and study dependence and independence notions in various areas such as quantum foundations and social choice theory \cite{abramsky2021team,PacuitY16}. Interestingly the novel characteristic  features of team semantics are also present in the independently developed  frameworks of inquisitive and separation logic \cite{DBLP:journals/jphil/CiardelliR11,1029817}. In fact, several translations between logics in these three differently motivated frameworks are now known 
(see, e.g.,  \cite{HAASE2021103063,DBLP:journals/ndjfl/CiardelliIY20}). 

The complexity aspects of logics in team semantics have been studied actively during the past years. The non-classical interpretation of disjunction in team semantics has the effect that, e.g., the model checking problem can become $\np$-complete for extremely simple quantifier-free formulas with disjunction and dependence atoms only (see Section~\ref{sec:Deriving tractable fragments}). 
An up-to-date study of the complexity of model checking for logics in team semantics with dependence, inclusion and independence atoms can be found in \cite{10.1145/3471618}.
On the other hand, the study of the  problems  of counting and enumerating  of teams satisfying a fixed formula has not yet been extensively studied (see \cite{haak10963,MeierR18,Haaketal20}).

The  complexity properties of some of the logics in team semantics are sensitive to subtle variations in the interpretation of disjunction and the existential quantifier. For example, under the nowadays standard so-called lax semantics inclusion logic is  equivalent to the positive greatest fixed point logic ($\PGFP$) and hence captures $\Ptime$ over finite (ordered) structures \cite{gallhella13}.  On the other hand, under the so-called strict semantics inclusion logic becomes equivalent to  existential second order logic ($\ESO$) and  captures $\np$ over finite structures \cite{galhankon13}.

\paragraph{Contributions of the paper.} While most of the logics in team semantics
can  in fact be translated into  $\ESO$, and hence to propositional satisfiability, the expressive power of such logics can vary greatly between $\fo$ and  $\ESO$ 
 but no modular approach for determining the expressivity of such a logic exists so far.
In this article, we develop a toolbox for translating efficiently algorithmic tasks such as model-checking, counting or enumeration for a team-based logic $\calL$ into the analogous problems for various fragments of $\pb{SAT}$. 
Our approach 
relies on the modularity in the very definition of team-based logics that allows to determine the relevant fragment of $\pb{SAT}$ by the selection of team connectives, operators, and atomic dependences allowed in the formulas of $L$. We then utilize our approach to obtain a wide range of new complexity results. In particular, in many cases, by using a finer-grained analysis of the context, the translation ends up in well-known or provable tractable fragments of $\pb{SAT}$ in $\nl$ or $\Ptime$. 

On the one hand, team-based logics are very versatile and powerful languages to capture many properties that arise in mathematics, statistics, computer science, databases, physics or social sciences. On the other hand, many highly efficient solvers exist for $\pb{sat}$ (for decision, counting but also enumeration). So, the existence of automatizable compilations  from the former to the latter that respect the intrinsic complexity of the problems is of true interest.

Some complexity results for team-based logics have been obtained in the past following this path.
For example, the model-checking problem for a fixed inclusion logic formula can be reduced in logarithmic space to the satisfiability problem of Dual-Horn clauses providing a direct proof for the result that the data complexity of  inclusion logic is in PTIME (see, e.g., \cite{10.1145/3471618}). This result  is analogous in spirit  to the well-known  results of Gr\"adel on the complexity of Horn and Krom fragments of second-order logic \cite{DBLP:journals/tcs/Gradel92}. It is worth to mention also that besides ad-hoc approaches, methods from  game semantics have also been used to derive complexity results for reasoning tasks in this area (see e.g. \cite{graedelH-2016,10.1145/3471618}) and connections with constraint satisfacion problems have been studied \cite{MR3566703}).

\paragraph{Organization of the paper.} Section~\ref{preliminaries} contains the relevant background on team-based logics and complexity theory including the introduction of new variants of the team semantics disjunction. It also presents a seemingly new tractable subclass of $\pb{SAT}$. Section~\ref{sec: translation to sat} contains the modular translations  from  the  model checking problems for formulas of various team-based logics to fragments of $\pb{SAT}$. Finally, Section~\ref{sec:Deriving tractable fragments} showcases the usefulness of the approach via the identification of a number of new fragments for which the aforementioned reasoning tasks are shown to be tractable/less complex  compared to all formulas of the logic.

\section{Preliminaries}\label{preliminaries}

\subsection{Syntax and semantics of team-based logics} We present here the basics of first-order team semantics.
 \begin{definition}
Let $\mA$ be a structure with domain $A$, and $V=\{x_1,\ldots,x_k\}$ be a finite (possibly empty) set  of
variables. Any finite set $X$ of assignments $s\colon V\rightarrow A$ is called a
  \emph{team} of $\mA$ with domain $\Dom(X) = V$.
\end{definition}
 For a tuple $\tu x=( x_1,\ldots,x_n)$, where  $x_i\in V$,  $X(\tu x):=\{s(\tu x) : s \in X\}$ is the $n$-ary relation of $A$, where $s(\tu x):=( s(x_1),\ldots,s(x_n))$.
\longversion{
 For $W\subseteq V$, we use $X \upharpoonright W$ to denote  the team obtained by restricting all assignments of $X$ to  $W$.
} 
  The set of free variables of a formula $\phi$, denoted $\Fr(\phi)$, is  defined as in first-order logic, taking into account that free variables may arise also from any atomic dependences.


\begin{definition}\label{defin:logicaloperations}
Let $\mA$ be a structure, $X$ be a team of $A$, and $\phi$ be a first-order formula such that $\Fr(\phi)\subseteq \Dom(X)$. 
\begin{description}
\item[lit:] For a first-order literal $\alpha$, $\mA \models_X \alpha$ iff for all $s \in X$, $\mA \models_s \alpha$.
\item[$\vee$:]  $\mA \models_X \psi \vee \theta$ iff  there are $Y$ and $Z$ such that $Y \cup Z=X$,  $\mA \models_Y \psi$ and $\mA \models_Z \theta$. 
\item[$\wedge$:] $\mA \models_X \psi \wedge \theta$ iff $\mA \models_X \psi$ and $\mA \models_X \theta$.
\item[$\exists$:]  $\mA \models_X \exists x \psi$ iff there exists a function $F : X \rightarrow \mathcal{P}(A)\setminus \{\emptyset\}$ s.t. $\mA \models_{X(F/x)} \psi$, where $X(F/x) =  \{s(m/x) : s \in X, m \in F(s)\}$ and $s(m/x)$ is the assignment that maps $x$ to $m$ and otherwise agrees with $s$.
\item[$\forall$:] $\mA \models_X \forall x \psi$ iff $\mA \models_{X(A/x)} \psi$, where $X(A/x) = \{s(m/x) : s \in X, m \in A\}$.
\end{description}
A sentence $\phi$ is  \emph{true} in $\mA$ (abbreviated $\mA \models \phi$) if $\mA \models_{\{\emptyset\}} \phi$. Sentences $\phi$ and $\phi'$ are  \emph{equivalent}, $\phi \equiv \phi'$, if for all models $\mA$, $\mA \models \phi \Leftrightarrow \mA \models \phi'$.
\end{definition}

The above is called the \emph{Lax Semantics}. In the \emph{Strict Semantics}, the semantic rule for disjunction is modified by adding the requirement $Y\cap Z = \emptyset$, and  the clause for the existential quantifier  is replaced by 
\begin{description} 
\item $\mA \models_X \exists x \psi$ iff there exists a function $H : X \rightarrow A$ s.t. $\mA \models_{X(H/x)} \psi$, with $X(H/x) = \{s(H(s)/x) : s \in X\}$. \end{description}
The meaning of first-order formulas is invariant under the choice between the strict and the lax semantics. 
First-order formulas satisfy what is known as the \emph{flatness} property:
 $\mA \models_X \phi$, iff  $\mA \models_s \phi$  for all $s \in X$.
Next we will give the semantic clauses for the new atomic dependences:

\begin{definition}\label{atoms}
Let $\tu x$, $\tu y$, $\tu z$, and $\tu u$ be tuples of variables such that $\tu x$ and $\tu y$ are of the same length. Also, let $v$ be any variable.
\begin{itemize}
\item  $\dep(\tu x, v)$ is a \emph{dependence atom}, with the semantic rule
\begin{description}
\item $\mA \models_X \dep(\tu x, v)$ iff for all $s, s' \in X$, if $s(\tu x)=s'(\tu x)$, then $s(v)=s'(v)$;
\end{description}
\item   $\indep{\tu x}{\tu u}{\tu z}$ is a \emph{conditional independence atom}, with the semantic rule
\begin{description}
\item $\mA \models_X {\tu x}\ \bot_{\tu u}\ {\tu z}$ iff for all $s, s' \in X$ such that $s(\tu u)=s'(\tu u)$, there exists an assignment $s'' \in X$ such that $s''(\tu x\tu u \tu z)=s(\tu x\tu u)s'(\tu z)$.
\end{description}
Furthermore, when $\tu z$ is empty, we write ${\tu x}\bot{\tu u}$ as a shorthand for $\indep{\tu x} {\tu z}  {\tu u}$, and  call it a \emph{pure independence atom};
\item $\tu x \subseteq \tu y$ 
is an \emph{inclusion atom}, with the semantic rule 
\begin{description}
\item $\mA \models_X \tu x \subseteq \tu y$ iff for all $s\in X$ there exists an $s' \in X$ such that $s'(\tu y)=s(\tu x)$. 
\end{description}
\item  $\tu x\ \Upsilon\ \tu y$ 
is an \emph{anonymity atom}, with the semantic rule 
\begin{description}
\item $\mA \models_X \tu x\ \Upsilon\ \tu y$ iff for all $s\in X$ there exists an $s' \in X$ such that $s'(\tu x)=s(\tu x)$ and
$s'(\tu y)\ne s(\tu y)$. 
\end{description}
\item $ \tu x \mid \tu y$ is an \emph{exclusion} atom, with the semantic rule 
\begin{description}
\item $\mA \models_X \tu x \mid \tu y$ if and only if $X(\tu x) \cap X(\tu y) = \emptyset$.
\end{description}

\item $\dep(v)$, a special case of the dependence atom, is a \emph{constancy atom}, with the semantic rule

\begin{description}
\item $\mA \models_X \dep(v)$ iff for all $s, s' \in X$, $s(v)=s'(v)$.
\end{description}
\end{itemize}

\end{definition}

The anonymity atom was introduced in \cite{paredaens89} with the name \emph{afunctional dependence}. It represents a strong opposite to functional dependence in database.

\begin{example}
In the context of databases with attributes that include $x,y,z$, dependence atom $\dep(x,y)$ says that attribute $y$ is functionally dependent on (or determined by) attribute $x$ as e.g. in $\dep(employee,boss)$. Independence atom $\indep{x}{y}{z}$ says that attributes $x$ and $z$ are independent of each other if $y$ is kept constant as e.g. in $salary\bot_{rank}gender$. Example~\ref{example: horizontal decomposition} demonstrates the use of a combination of dependence and anonymity atoms in a database. 
Example~\ref{example: translation of cycle formula} contains an application of the inclusion atom to express the property of a graph of containing a cycle.
\end{example}

\begin{definition}

Let $\scrS\subseteq\{\dep(\cdot,\cdot), \indep{}{}{},\subseteq, \anon{}{}, \const\}$. The formulas of  $\fo(\scrS)$ are obtained by extending the syntax of $\fo$ by atoms built from the operators in $\scrS$. Its semantics is obtained by extending Definition~\ref{defin:logicaloperations} by the semantic rules of Definition~\ref{atoms} for each operator in $\scrS$.
\end{definition}

As already mentioned, several of the above logics have received a lot of interest over the last 15 years. The fragment $\FOdep$, also denoted $\df$, is called dependence logic. Independence logic $\indlogic$, inclusion logic $\FOinclusion$, and anonymity logic $\FOanon$ are defined analogously using independence, inclusion, and anonymity atoms, respectively, only.
It is easy to see that all these fragments are indeed expressible in existential second order logic. 
There are some basic translations between them:
$\df$ and $\FOinclusion$ can be translated into $\indlogic$ but not to each other, while $\FOanon$ and $\FOinclusion$ can be translated to each other. Furthermore, conditional independence atoms can be simulated by formulas using only pure independence atoms. All these translations require the  use of propositional connectives and quantifiers. In the current paper we are concerned with weak fragments of these and related logics, composed of some atoms as in Definition~\ref{atoms} by means of very limited applications of conjunction, disjunction and the quantifiers.

It is easy to see that the flatness property is lost immediately when $\fo$ is extended by any of the atoms defined above. On the other hand,  it is straightforward to check that all $\df$-formulas satisfy the following  strong \emph{downwards closure} property:  if $\mA \models_X \phi$ and   $Y\subseteq X$, then $\mA \models_Y \phi$. Furthermore, Anonymity and inclusion logics are closed under unions i.e. the union of any set of teams satisfying a given formula of these  logics itself satisfies the formula (\cite[Theorem 4.9]{galliani12}).  

Another basic property of formulas and logics in team semantics is called \emph{locality} meaning:

 $$\mA \models_X \phi \ \mbox{ iff }   \ \mA \models_{X \upharpoonright \Fr(\phi)} \phi.$$
Under the lax semantics all of the operators and atoms defined above preserve locality when added to first-order logic whereas for the strict semantics locality only holds for formulas of dependence logic 
(see 
\cite{galliani12} for details).

\subsubsection{Block disjunctions}

In team semantics, disjunction splits a team into two subteams each satisfying separate properties. In many contexts, it may  however be interesting to be able to express the splitting relative to values of a given variable or set of variables. This is the case for example in databases where inconsistent data needs to be repaired by splitting rows between the ones that satisfy a given dependence constraint on a set of attributes and those that do not satisfy it (see Example~\ref{example: horizontal decomposition} below). This feature can be very simply expressed in team semantics through the introduction of a new  variant of disjunction. If $X$ is a team, then $Y\subseteq X$ is an 
$x$-\emph{block},
   if ($s\in Y\ \&\ s'\in X\ \&\ s(x)=s'(x)$) imply $s'\in Y$.

\begin{definition}[Block Disjunction]\label{def:blockdis}
Let $X$ be a team and $x\in Dom(X)$.
Define the \emph{block disjunction}, $\vee^b_x$, by the clause: \\  
$X \models \phi \vee^b_x \psi$ iff
there exists $Y, Z$ such that

\begin{enumerate}
\item $X=Y\cup Z$, $Y \models \phi$ ,  $Z\models \psi$
   \item  $Y$ and $Z$ are $x$-blocks.
\end{enumerate}

We obtain the \emph{strict block disjunction}, $\vee_x$, by adding the further condition
\begin{enumerate}
\item[3.]          $Y(x) \cap Z(x) = \emptyset$ 
\end{enumerate}
\end{definition}

Note that the strict block disjunction is a strictly stronger constraint than the usual strict disjunction as 3. above implies $Y\cap Z=\emptyset$ but the converse is not true. Interestingly the strict block disjunction is similar to the separating conjunction of separation logic (see \cite{HAASE2021103063}).

\begin{example}~\label{example: horizontal decomposition}
In databases, horizontal decomposition amounts to splitting a relation $R$ into two parts: one part, say $R_1$, satisfying a given functional dependency $x\rightarrow y$, the other one, $R_2$ made of tuples that do not satisfy the dependency or, in other words that satisfy the afunctional dependency $x \nrightarrow y$. Note that, in this approach, it is expected that two tuples $t_1\in R_1$ and $t_2\in R_2$ can not have the same value for $x$ i.e. $t_1(x)\neq t_2(x)$ (see the notion of  an $X$-complete set in~\cite{DeBraThesis87,DBLP:journals/ipl/BraP83}). In team semantics, such a decomposition  can be expressed by the following formula:
\[
\dep(x,y) \vee_x^b \anon{x}{y}.
\]
\end{example}
It turns out that allowing block disjunction will not change the complexity of model checking for several tractable fragments hence reasoning about data decomposition becomes tractable in such contexts.

\longversion{
\begin{lemma}TFAE for downward closed $\phi$ and $\psi$:
\begin{enumerate}
    \item $X\models\phi\vee^b_x\psi$
    \item $X\models\exists u\exists v\exists w(\dep(x,u)\wedge\dep(x,v)
    \wedge\dep(x,w)\wedge (u\ne v\vee\phi)\wedge (u\ne w\vee\psi)\wedge(u=v\vee u=w))$.
\end{enumerate}

\end{lemma}

\begin{proof}
Suppose first $X\models \phi\vee^b_x\psi$. Let $X=Y\cup Z$ as in Definition~\ref{def:blockdis}. Let us extend $X$ to $X'$ (and respectively $Y',Z'$) by giving values to $u,v$ and $w$ as follows: Let $a\ne b$. Suppose $s\in X$. Let $s(u)\equiv a$ and $s(v)=a$ except $s(v)=b$ if $s\in Z\setminus Y$. Let also $s(w)=a$ except $s(w)=b$ if $s\in Y\setminus Z$. Of course, $X'\models\dep(x,u)$. To prove $X'\models \dep(x,v)$ let us assume $s(x)=s'(x)$, where $s,s'\in X'$. Then $s$ and $s'$ are in the same $x$-block. If they are both in $Y$, then $s(v)=s'(v)=a$. If they are both in $Z\setminus Y$, then $s(v)=s'(v)=b$. Similarly, $X'\models\dep(x,w)$. To prove $X'\models u\ne v\vee\phi$ we note that $X'=(Z'\setminus Y')\cup Y'$, $Z'\setminus Y'\models u\ne v$ and $Z'\models\phi$.  Similarly, $X'\models u\ne w\vee\psi$. Finally, $X'\models u=v\vee u=w$ by construction.

Suppose then an extension $X'$ of $X$ exists obtained by giving values to $u,v,w$ such that 
\begin{multline*}
X'\models\dep(x,u)\wedge\dep(x,v)
    \wedge\dep(x,w)\wedge \\ (u\ne v\vee\phi)\wedge (u\ne w\vee\psi)\wedge(u=v\vee u=w).  
\end{multline*}

    Let $$Y=\{s\restriction \dom(X) : s\in X',  s(u)=s(v)\}$$
    $$Z=\{s\restriction \dom(X) : s\in X',  s(u)=s(w)\}.$$ We show that $Y$ is an $x$-block. Suppose therefore $s\restriction \dom(X)\in Y$ and $s(x)=s'(x)$, where $s,s'\in X'$. Since $X'\models\dep(x,u)\wedge\dep(x,v)$ and  $s(x)=s'(x)$, also $s(u)=s'(u)$ and $s(v)=s'(v)$. Since $s\restriction \dom(X)\in Y$, $s(u)=s(v)$, whence $s'(u)=s'(v)$ and hence $s'\in Y$. Similarly, $Z$ is an $x$-block. Since $X'\models u\ne v\vee \phi$, $X'=X'_1\cup X'_2$, where $X'_1\models u\ne v$ and $X'_2\models\phi$. Clearly,  $Y\subseteq X'_2$, whence $Y\models\phi$.  Similarly, $Z\models\psi$.
\end{proof}

The above Lemma holds for the strict block disjunction, too, if we add $v\ne w$ as a conjunct in 2.
In consequence, the block disjunction and its strict variant both preserve downward closure and locality.
} 

\subsection{Complexity classes and problems}

We suppose the reader is familiar with basic complexity classes such as $\logspace$ (logarithmic space), $\nl$ (non deterministic logarithmic space), $\Ptime$ (polynomial time), $\np$ (non deterministic polynomial time). In all the paper, $\calL\subseteq \calC$ will mean:  for any fixed formula $\varphi\in \calL$, the model-checking problem for $\varphi$ is in $\calC$. I.e. we are in the paradigm of \textit{data complexity}

\subsubsection{Counting and enumeration problems}

\newcommand\classFont[1]{\textnormal{#1}}
\renewcommand{\P}{\protect\ensuremath{\classFont{P}}\xspace}
\newcommand{\NP}{\protect\ensuremath{\classFont{NP}}\xspace}
\newcommand{\cP}{\protect\ensuremath{{\classFont{\#P}}}\xspace}
\newcommand{\totP}{\protect\ensuremath{{\classFont{TotP}}}\xspace}
\newcommand{\fP}{\protect\ensuremath{{\classFont{FP}}}\xspace}
\newcommand{\PP}{\protect\ensuremath{\classFont{PP}}\xspace}
\newcommand{\coNP}{\protect\ensuremath{\classFont{coNP}}\xspace}
\newcommand{\cTeam}[1]{\protect\ensuremath{\#{#1}^{\textnormal{team}}}\xspace}
\newcommand{\cTeamne}[1]{\protect\ensuremath{\#{#1}^{\textnormal{team*}}}\xspace}
\newcommand{\calA}{\mathcal{A}}


We quickly recall some complexity classes for counting problems that are relevant for team-based logics.
For a complexity class $\calC$, the counting analogue of $\calC$ can be defined as follows.
Let $L\subseteq \{0,1\}^*\times \{0,1\}^*$. For $x\in \{0,1\}^*$, one denotes by $L(x)=\{y: (x,y)\in L\}$. The relation $L$ is said to be polynomially balanced if for all $x\in \{0,1\}^*$ and $y\in L(x)$, $|y|$ is bounded by a polynomial in $|x|$. 

\begin{definition}\label{def: counting by witness predicate} Let $\calC$ be a complexity class. 
  A function $f \colon \{0,1\}^* \rightarrow \mathbb{N}$ is in $\#\cdot \calC$ if there is a polynomially balanced $L \in \calC$ such that for all $x \in \{0,1\}^*$:
  \[	f(x) = |L(x)|.\]
\end{definition}
Obviously $\cP= \#\cdot \P$, where $\cP$ refers to the class defined in the usual way via non deterministic machines, and it is known that $\cP\subseteq \#\cdot  \NP \subseteq \#\cdot  \coNP = \cP^{\NP}$, where, under reasonable complexity-theoretic assumptions, all these inclusions are strict; see, e.g.,  \cite{HemaspaandraV95}. 
In order to connect the above function classes to logics in team semantics, we  define $\cTeam{L}$ to consist of functions counting non-empty satisfying teams for \mbox{$L$-formulas}. 

\begin{definition}
	For a logic $L$, $\cTeam{L}$ is the class of all functions $f \colon \{0,1\}^* \to \mathbb{N}$ for which there is a vocabulary $\sigma$ and an $L$-formula $\phi(\tu x)$ over $\sigma$ with a tuple $\tu x$ of free first-order variables such that for all $\sigma$-structures $\calA$ with a built-in linear order $\leq$, 
\[f((\calA)) = |\{X \in \textrm{team}(\calA, (\tu x)) : X\neq \emptyset \textrm{ and } \calA \models_X \phi(\tu x)\}|.\]	
Above, the input to $f$ is strictly speaking an binary encoding of $\calA$. The binary inputs that do not encode some $\sigma$-structure as above are mapped to 0. 
\end{definition}
It was shown in \cite{haak10963}, e.g.,  that $\cTeam{\indlogic}= \#\cdot  \NP$ and that 
$\cTeam{\indlogic}\subseteq \cP$. Furthermore, dependence logic was shown to define a $\#\cdot  \NP$-complete but it remains open whether it is equal to
 $\#\cdot  \NP$.
 
 In order to state some results regarding enumeration problems, we recall the following definition.
  \begin{definition}
 Let $L\subseteq \{0,1\}^*$ be a polynomially balanced predicate such that $L\in \Ptime$. Problem $L$ is enumerable in polynomial delay and polynomial space if there is an algorithm which upon input $x\in \{0,1\}^*$, output elements of the  set $L(x)$ one by one, without repetition with a polynomial in $|x|$ delay between two elements and polynomial space. 
 \end{definition}

\subsection{The complexity of satisfiability problems}

\newcommand{\equivDH}{\mathsf{EDualHorn} }
\newcommand{\oneequivDH}{\mathsf{1\!-\!EDH} }
\newcommand{\leftC}[1]{\mathsf{Left}(#1)}
\newcommand{\rightC}[1]{\mathsf{Right}(#1)}
\newcommand{\clause}[1]{\mathsf{clause}(#1)}
\renewcommand{\DH}{\textsc{DH}}
\newcommand{\EQ}{\textsc{EQ}}
\newcommand{\Upos}{\textsc{U}^+}
\newcommand{\Uminus}{\textsc{U}^-}
\newcommand{\propagate}[1]{\mathsf{Propagate}(#1)}
\newcommand{\coherency}[1]{\mathsf{Coherency}(#1)}

A  propositional formula $\varphi$  in conjunctive normal form is called Horn (resp. dual-Horn) if each of its clauses contain at most one positive (resp. negative) literal. It is $2$-CNF if all its clauses have at most two literals and Implicative $2$-CNF if it is both $2$-CNF and (dual) Horn. A clause is a unit positive (resp. negative) clause if it is restricted to one (resp. negated) variable only. We will consider propositional formulas with possibly existentially quantified variables. One denotes by $\var{\varphi}$ (resp. $\Fr{(\varphi)}$) the set of (resp. free) variables of $\varphi$.
 
It is well-known that $k$-$\pb{SAT}$, the satisfiability problem for $k$-\pb{cnf} formulas, is $\np$-complete for $k\geq 3$ and $\nl$-complete for $k=2$. For of Horn or  dual-Horn formulas the satisfiability problem is $\Ptime$-complete. Recall also that dual-Horn (resp. Horn) formulas are closed under union (resp. intersection) of models and that one can compute the maximal (resp. minimal) model for inclusion in $\Ptime$. We introduce below a new fragment of $\pb{SAT}$ and prove it is tractable.


\begin{definition}[$\oneequivDH$]
A formula $\varphi$ is $\oneequivDH$ if it is a conjunction of clauses of the following kinds:

\begin{itemize}
\item $\Upos$ (resp. $\Uminus$) : unit  (resp. negative) clauses

\item  $\DH$: dual-Horn clauses 
\[
\begin{array}{cc}
C\equiv x \rightarrow y_1 \vee y_2\vee \dots \vee y_k    & \mbox{ with } k\geq 1 
\end{array}
\]

with the additional property that each 
$x\in \var{\varphi}$ appears at most once as the left hand part of a clause. \longversion{For such a clause, one denotes 
$\leftC{C}=x$ and $\rightC{C}=\{y_1,...,y_k\}$.}

\item $\EQ$: Implicative $2$-CNF clauses of the form

\[
y_1\rightarrow x, y_2\rightarrow x, \ldots, y_k\rightarrow x
\]
for some $C\equiv x \rightarrow y_1 \vee y_2\vee \dots \vee y_k$ belonging to $\DH$. Again, each variable $y$ can appear at most once in the left hand part of a clause in $\EQ$. \longversion{Such clauses are called \textit{back-implications}.}
\end{itemize}
\end{definition}

Alternatively, non unit clauses of formulas in $\oneequivDH$ of the form (left and right are chosen arbitrarily when $k=1$)

\longversion{
\[
\begin{array}{cc}
C\equiv x \rightarrow y_1 \vee y_2\vee \dots \vee y_k    & \mbox{ with } k\geq 1 \\
C\equiv x \leftrightarrow y_1 \vee y_2\vee \dots \vee y_k    &  \mbox{ with } k\geq 1
\end{array}
\]
}

\shortversion{
\[
\begin{array}{rcl}
x \rightarrow y_1 \vee y_2\vee \dots \vee y_k    & \mbox{ or } & x \leftrightarrow y_1 \vee y_2\vee \dots \vee y_k 
\end{array}
\]
}

\noindent with the additional global condition  that: a variable $x$ can appear in the left hand side of such clauses only once; a variable $y_i$ can appear only once in the right hand side of an equivalence clause. The number of occurrences of a variable in the right hand side of a pure implication clause (DH) is not limited. An example of a $\oneequivDH$ formula is given in Example~\ref{example: translation of cycle formula}. 
We prove the following.

\begin{proposition}\label{prop: EDH in NL}
The satisfiability problem for formulas in $\oneequivDH$ is in $\nl$.
\end{proposition}
 
\longversion{ 
Before proving the proposition, let us fix some notations. Let $\varphi$ in $\oneequivDH$ and $y\in \var{\varphi}$. One defines $\propagate{y,\varphi}$ as follows:
\begin{itemize}
    \item $y\in \propagate{y,\varphi}$
    \item if there exists $C\in \EQ$ s.t. $\leftC{C}\in \propagate{y,\varphi}$ then $\rightC{C}\subseteq \propagate{y,\varphi}$
\end{itemize}

Note that since $C\in \EQ$ above, $\rightC{C}$ is restricted to one element. The set $\propagate{y,\varphi}$ will contain all variables whose value is forced to true when $y$ is set to true. This gives  rise to the following definition.

\begin{definition}[Safe variable] Let $\varphi$ in $\oneequivDH$ and $y\in \var{\varphi}$. Variable $y$ is \textit{safe} if:
\[\propagate{y,\varphi}\cap \Uminus = \emptyset.\]
\end{definition}

\begin{proof}[Proof of Proposition~\ref{prop: EDH in NL}]
The decision procedure is described in Algorithm~\ref{algo EDH} and~\ref{algo coherency}. The principles mix some reachability techniques with unit propagation.
The rough idea is as follows. We consider in turn all variables  appearing in a unit positive clause. Let $x$ be such a variable. Let $C$ be the (at most unique) clause $C\equiv x \rightarrow y_1 \vee y_2\vee \dots \vee y_k$ that contain $x$ as head. Then, one try to guess a variable $y_i$ that could be set to $1$. For this, $y_i$ must not be a negative unit clause and, more generally, must be safe. If such a $y_i$ exists, one can immediately stop if it belongs to a positive unit clause. If not, one continue the process with $y_i$ instead of $x$. After at most $|\var{\varphi}|$ steps, if the algorithm has not stopped before, we know that a variable has been seen at least twice and that no contradiction emerges from setting $x$ to $true$. If this process concludes positively for all unit clauses, then one can deduce $\varphi$ is satisfiable.    

At each step one needs, in the worst case, to guess a index of a variable ($y_i$) and check if it is safe. The guessing requires a logarithmic number of bits. 
Checking if a variable is safe can be done in logarithmic space because of the one occurrence condition of all variables appearing as head of a back-implication.

\begin{algorithm}[t]
  \caption{Decision procedure for $\oneequivDH$ formulas}~\label{algo EDH}
  \begin{algorithmic}[1]
  \State \textbf{Input:} $\varphi\in \oneequivDH$
  \For{$x\in \Upos$}
  \If{$\propagate{x,\varphi}\cap \Uminus \neq \emptyset$} $\return$ $\reject$ \EndIf
  \If{$\coherency{x,\varphi, |\var{\varphi}|}=\false$} $\return$ $\reject$
  \EndIf
  \EndFor
  \State $\return$ $\accept$
  \end{algorithmic}
  \end{algorithm}

  \begin{algorithm}[t]
  \caption{$\coherency{x,\varphi, \alpha}$}~\label{algo coherency}
  \begin{algorithmic}[1]
  \If{$\alpha=0$} $\return$ $\accept$ \EndIf
  \If{$\forall C\in \DH \ x\neq \leftC{C}$} $\return$ $\accept$
  \EndIf
  \State Let $C\in \DH$ s.t. $x=\leftC{C}$ 
  \State $\guess$ $y\in \rightC{C}$
  \If{$y\in \Upos$} $\return$ $\accept$ \EndIf
  \If{$y$ is safe} $\return$ $\coherency{y,\varphi, \alpha-1}$ \Else \ $\return$ $\reject$ \EndIf
  \end{algorithmic}
  \end{algorithm}

First suppose $\varphi$ is satisfiable and let $I$ be a satisfying assignment. Obviously, for all $x\in \Upos$, $I(x)=1$ and, by construction, for all $y\in \propagate{x,\varphi}$, it must also holds that $I(y)=1$. Hence, $x$ is safe and, for such $x$ a reject answer can only come from the call of $\coherency{x,\varphi, |\var{\varphi}|}$. For a variable $v$, let us call $C_v$, when it exists, the only clause such that $v=\leftC{C_v}$. Now let $x=y_0,...,y_t$, $t\leq |\var{\varphi}|$ be the longest sequence such that, for all $i< t$: 

\begin{itemize}
    \item $C_{y_i}$ exists 
    \item $y_{i+1}\in \rightC{C_{y_i}}$ and $y_{i+1}\not\in \Upos$
    \item $I(y_{i+1})=1$
\end{itemize}

Since we are considering a sequence of implications starting from $X\in \Upos$, there must exists a sequence satisfying also the last item i.e. with  $y_{i+1}$ that is set to $1$ by $I$. The $y_i$ well correspond to a list of consecutive calls of $\coherency{y_i,\varphi, \alpha}$ without rejecting. Since $t$ is maximal, we are in one of the following cases:

\begin{itemize}
    \item Assume $t=|\var{\varphi}|$. In that case, the algorithm, when called with $y_t$ accept
    \item If this is not the case, then it might be that $C_{y_t}$ does not exists. In that case, the algorithm returns true also
    \item If $C_{y_t}$ exists, if there exists $y_{t+1}\in \rightC{C_{y_t}}$ such that $y_{t+1}\in \Upos$ then again the algo accepts
    \item Finally, it might be that for all possible choices of $y_{t+1}\in \rightC{C_{y_t}}$, $y_{t+1}\not\in \Upos$ and  $I(y_{t+1})=0$. But this clearly contradicts the fact that $I\models \varphi$.
\end{itemize}

So, for all $x\in \Upos$, we have exhibited a sequence of possible non deterministic choices that lead to acceptation. Hence Algorithm~\ref{algo EDH} returns true on input $\varphi$.

Suppose now that a formula $\varphi$ is accepted by Algorithm~\ref{algo EDH}.
Then, for all $x\in\Upos$, $x$ is safe and $\coherency{x,\varphi, |\var{\varphi}|}$ outputs true. Let
\begin{multline*}
 V_x = \propagate{x,\varphi} \cup  \{y: y \mbox{ is guessed during }\\ \mbox{ an accepting run of } \coherency{x,\varphi, |\var{\varphi}|}\}   
\end{multline*}
Define $V^+=\bigcup_{x\in\Upos}V_x$ and 
$I^+:V^+\rightarrow\{0,1\}$ by $I^+(v)=1$, for all $v\in V^+$. Note that two sets $V_x$, $V_{x'}$ may intersect but this is coherent with $I^+$ that assigns everything to $1$. Let now, $V^-=\var{\varphi}\backslash V^+$. In particular $\Uminus\subseteq V^-$. Let us define $I^-$ by, for all $v\in V^-$, $I^-(v)=0$. It is easily seen that $I:V\rightarrow\{0,1\}$ defined by $I(v)=I^+(v)=1$ for $v\in V^+$ and  $I(v)=I^-(v)=0$ for $v\in V^-$ satisfies $\varphi$.\end{proof}
} 

\section{Modular translations from team-based logics to \pb{SAT}}~\label{sec: translation to sat}

We describe in this section a modular approach to translate  model checking problems and more general reasoning tasks for logics in team semantics to Boolean satisfiability  problems. 

For a team-based formula $\varphi$, let $\subformula{\varphi}=\{\varphi_0,\varphi_1,\varphi_2,\ldots, \varphi_t\}$, with $\varphi_0=\varphi$,  the set of subformulas of $\varphi$ 
\longversion{(with $i<j$ when $\varphi_j$ is a subformula of $\varphi_i$)}.
Let $\mX$ be the following set of propositional variables $\mX=\{X[s]: s\in A^{r_0}\}$.

We prove the following general result whose significance in terms of computational bounds for various fragments will appear clearly in the rest of the paper. In particular, it will enable us to identify new tractable fragments of team-based logics that were not known so far.

\begin{theorem}~\label{theorem: translation}
Let $\scrS\subseteq \{\dep(\cdot,\cdot), \bot,\subseteq, \anon{}{}, \const, \mid\}$ and $\varphi\in\FO{\scrS}$. There is a logspace algorithm which on input $\mA$ and $r\in \mathbb{N}$ with $r\geq |\Fr{(\varphi})|$, outputs a propositional formula $\Psi$ 
with $\mX=\Fr{(\Psi)}\subseteq \var{\Psi}$ such that, for all team $X\subseteq A^r$, the following are equivalent:
\begin{itemize}
    \item $\mA\models_X\varphi$
    \item There is a assignment $I:\mX\longrightarrow \{0,1\}$ that satisfies $\Psi$ such that for all $X[s]\in \mX$:  $I(X[s])=1 \iff s\in X$.
\end{itemize}

Moreover, $\Psi$ is of the form:

\[
\Psi=\exists p_1\ldots \exists p_n\bigwedge_{\varphi_i\in \subformula{\varphi}} F_i
\]

\noindent where $\{p_1,...,p_n\}=\var{\Psi}\backslash \mX$, and each $F_i$ depends on the subformula $\varphi_i$ and is of one of the types described in Figure~\ref{fig:recap_table}.
\end{theorem}

Let us comment on Figure~\ref{fig:recap_table}. Line entries must be understood as follow: if a subformula is of a given form then the clauses that are produced during the translation can take the syntactic forms described by columns where a $\star$ is found  for that line. For example, 
\longversion{ for line~\ref{rule:vee general+fo},}
if the current subformula $\varphi_i$ is of the form $\varphi_j\vee \varphi_h$ with $\varphi_j\in \fo$ then the propositional subformula $F_i$ is implicative 2-CNF (which implies Horn, dual-Horn etc). Sometimes, the fact that evaluation is made on a team which is itself $\logspace$ computable is isolated
\longversion{(see, for example line~\ref{rule:vee logspace+fo+downward})}. Alternatively, each column of the table indicates those connectives, quantifiers and atoms for which the given clause type suffices for the translation. For example, starting with formulas of a logic $\mathcal{L}$ that includes classical first-order logic, that uses dependence atoms $\dep(\cdot,\cdot)$ as only atoms and that allows universal quantification, conjunction, restricted used of disjunction between formulas of $\mathcal{L}$ and  first-order formulas, one ends up with propositional formulas that are always Horn.

The theorem is proved in a modular way that allow to control the final shape of the target propositional formula depending on the restricted set of atoms that are really used in  $\varphi$. We illustrate the method on the following example.
 
\begin{example}~\label{example: translation of cycle formula}
Let us consider formula $\varphi=\exists x\exists y (y\subseteq x \wedge E(x,y))$. It was shown in \cite{gallhella13} that sentence $\varphi$ is true in a graph $G=(A,E)$ and the empty team $X$ if and only if $E$ contains a cycle. 

Along the syntactic tree of $\varphi$, we construct inductively a set of clauses $\calC$ over a set of propositional variables $\mX$. Calling $\Psi$ the conjunction of clauses in $\calC$, we will have $G\models_X \varphi$ iff there exists an assignment of variables in $\mX$ such that $\Psi$ is satisfiable. 

One starts with $\mX=\calC=\emptyset$.
Let's first consider $\varphi=\exists x\varphi_1$ with $\varphi_1(x)= \exists y (y\subseteq x \wedge E(x,y)$. One adds the existentially quantified propositional variables, $X_1[s]$ for $s\in A$ and add to $\calC$ the following set  clause that, in accordance to the semantics of the existential quantifier, requires that at least one of the variables above is true:
\(
\bigvee_{s\in A} X_1[s].
\)
We now have to proceed with $\varphi_1(x)=\exists y \varphi_2(x,y)$ with $\varphi_2(x,y)= y\subseteq x \wedge E(x,y)$. One adds the existentially quantified propositional variables, $X_2[s]$ for $s\in A^2$ and add in $\calC$ as clauses, constraints that relate variables of $X_1$ type to variables of $X_2$ type. 

\[
\bigwedge_{s\in A}	X_1[s]\leftrightarrow \bigvee_{
		s'=(s,a),\  a\in A} X_2[s']	
\]

Let now $\varphi_3(x,y)=x\subseteq y$ and $\varphi_4(x,y)=E(x,y)$. To treat  $\varphi_3(x,y)$, no additional propositional variables are necessary since the arity of the team is not growing but one has to add a set of clauses that express  the meaning of the inclusion atom: 

\[
\bigwedge_{s\in A^2}	X_2[s]\rightarrow \bigvee_{s'\in A^2, \restrict{s'}{\tu y}=\restrict{s}{\tu x}} X_2[s'] 	
\]

Finally, to handle the information carried by $\varphi_4(x,y)$, one adds:

\[
\bigwedge_{s: s=(s(x),s(y))\in E} X_2[s] \wedge \bigwedge_{s: s=(s(x),s(y))\not\in E} \neg X_2[s] 	
\]

It is not hard to see that not only $G\models_X \varphi$ iff $\Psi$ is satisfiable but also that

\begin{itemize}
	\item $\calC$ is computable in $\logspace$
	\item The formula $\Psi$ obtained by conjunction of the clauses above is in $\oneequivDH$.
\end{itemize}

Hence the compilation ended in a fragment of $\pb{SAT}$ that is $\nl$ decidable. Note that this  matches exactly the complexity of the problem of finding a cycle in a graph that is expressed by the example sentence $\varphi$. 
\end{example}

 \newcommand{\logspaceshort}{\mathbf{L}}
 \begin{figure}[t]
     \centering
     \scalebox{0.95}{
\longversion{
\begin{tabular}{|c|c|c|c|c|c|c|c|}
\hline
    rule      & formula                  & U & I2C & 2C & H & DH & DH$_1$ \\ \hline\hline
$(\fo)$      &   $\varphi$:$\fo$           & $\star$  & $\star$ & $\star$  & $\star$ & $\star$  & $\star$     \\ \hline
$(\exists)$  &  $\exists x\varphi$                        &      &          &      &      & $\star$        & $\star$    \\ \hline
$(\forall)$  & $\forall x\varphi$                         &      & $\star$        & $\star$    & $\star$    & $\star$        &      \\ \hline
\ref{rule: wedge}  & $\phi\wedge\varphi$                         &      & $\star$        & $\star$    & $\star$    & $\star$        &      \\ \hline
\ref{rule: wedge FO} & $\fo\wedge\varphi$         &      & $\star$        & $\star$    & $\star$    & $\star$        & $\star$    \\ \hline
\ref{rule:general or} & $\phi\vee\varphi$        &      &          &      &      & $\star$        & $\star$    \\ \hline
\ref{rule:vee general+fo}& $\fo\vee\varphi$          &      & $\star$        & $\star$    & $\star$    & $\star$        & $\star$    \\ \hline
\ref{rule:vee general+logspace} & $\logspaceshort$, $\phi\vee\varphi$             &      &          & $\star$    &      & $\star$        & $\star$    \\ \hline
\ref{rule:vee logspace+fo+downward} & $\logspaceshort$, $\phi$:$\fo$, $\varphi$:dw & $\star$ & $\star$ & $\star$ & $\star$ & $\star$ & $\star$   \\ \hline
\ref{rule: or relativized block general} &   $\phi\vee_x^b\varphi$                       &      &          &      &      & $\star$        &      \\ \hline
\ref{rule: or relativized block logspace} &  $\logspaceshort$, $\phi\vee_x^b\varphi$                       &      &          &   $\star$   &      & $\star$        &      \\ \hline
(incl)       &   $\tu x\subseteq \tu y$                 &      &          &      &      & $\star$        & $\star$    \\ \hline
(anon)       &   $\anon{\tu x}{\tu y}$                       &      &          &      &      & $\star$        & $\star$    \\ \hline
(dep)        &    $\dep(\tu x,y)$                      &      &          & $\star$    & $\star$    &          &      \\ \hline
(indep)      & $\indep{\tu x}{\tu y}{\tu z}$                         &      &          &     &     &          &      \\ \hline
(excl)       &  $\tu x | \tu y$                      &      &          & $\star$    & $\star$    &          &      \\ \hline
(const)       &  $\dep(x)$ &   \multicolumn{6}{c|}{polynomial size DNF formula}     \\ \hline
\end{tabular}
} 

}
     \caption{Form of clauses after compilation to $\mathsf{SAT}$. $\logspaceshort$ means that the team $X$ is in $\logspace$. U is for unit clauses, I2C for Implicative-$2$CNF, 2C for $2$CNF, H for Horn, DH for Dual Horn, DH$_1$ for $\oneequivDH$. The last line means that the formula produced when translating a constancy atom is DNF.}
     \label{fig:recap_table}
 \end{figure}

\subsection{The proof of Theorem~\ref{theorem: translation}}

We now proceed with the complete description of the method. First, let us remark that for any fixed formula $\varphi\in \fo$, there exists a logarithmic space algorithm which, upon input $\mA$ of domain $A$, computes the maximal team $X\subseteq A^r$ with $r=|\Fr{(\varphi)}|$ such that:

\[\mA\models_X  \varphi.\]

Even more evidently, the complementation $X[A/x]$ of any team $X$ is computable in logarithmic space.

Let $\varphi$ be a formula of team logic, $\mA$ be a structure of domain $A$, and $X$ be a team with $\Fr{(\varphi)}\subseteq \dom(X)$. Let $r_0\geq |\dom(X)|$. We consider the set $\mX$ of propositional variables $X[s]$ for $s\in A^{r_0}$.
Starting from $\varphi$, $\mA$ and a possible team $X$, the evaluation of whether $\mA\models_X\varphi$ is obtained inductively by evaluating the truth of  subformulas  $\varphi_0=\varphi,\varphi_1,\varphi_2,\ldots, \varphi_t$ (with $i<j$ when $\varphi_j$ is a subformula of $\varphi_i$) of $\varphi$ over potential teams $X_0(=X), X_1,X_2,\ldots, X_t$ built from $X$.  This evaluation process will be controlled step by step by the introduction of Boolean clauses 
built over some existentially quantified propositional variables issued from $X_1,X_2,\ldots, X_t$. At each step, we will make clear which type of clauses are introduced.
Let $\calS=\{(\varphi,X,r_0)\}$.

\medskip 

\noindent \textbf{Initialization. } The clauses below will be needed only when one needs to provide information about a reference team inside the constraints.

\begin{enumerate}[label=\textbf{(I.\arabic*)}]
	\item\label{rule:init} In the general case, let $\calC=\{X[s] : s\in X\} \cup \{\neg X[s] : s\not\in X\}$.
	\item\label{rule:init union} In the special case where  $\varphi$ is closed under union, we set $\calC=\{\neg X[s] : s\not\in X\}$. 
\end{enumerate}

\textsf{Type of clauses:} all clauses that are introduced are unitary clauses.
\medskip

\noindent The propositional formula $\Psi$ is now constructed inductively as follows.
 As long as $\calS\neq \emptyset$, we apply the following rule:  Pick $(\varphi_i,X_i,r_i)$, $i\leq t$, in $\calS$ and apply the following rules.

 \medskip
 
\noindent \textbf{First-order rules.}

\begin{description}
%
%
  	
\item[($\fo$)] If $\varphi_i\in \fo$ then: $\calS:=\calS\backslash \{(\varphi_i,X_i,r_i)\}$  and $\calC:=\calC \cup \{\neg X_i[s]: \mbox{ for all } s\in A^{r_i} \mbox{ s.t. } \mA\not\models_{s}\varphi_i \}$.
 
Clearly : $\mA\models_{X_i}\varphi_i$ iff for all $s\in X_i$, $\mA\models_{s}\varphi_i$. 
 
\textsf{Type of clauses:} all clauses that are introduced are unitary. Note that this rule preserves $\logspace$ computability of the teams: if $X_i$ is $\logspace$ computable then so is $X_j$.

\item[($\exists$)] If $\varphi_i$ is  $\exists x \varphi_j$, then: $\calS := ( \calS \backslash \{(\varphi_i,X_i,r_i)\} ) \cup \{(\varphi_j,X_j, r_j=r_i+1)\}$ and 
  	\[
  		\textstyle
  	    	\calC:=\calC \cup \{X_i[s]\leftrightarrow \bigvee_{
  	  	  	s'=(s,a),\  a\in A} X_j[s']: s\in A^{r_i} \}, 
  	  	\]
%
%
%
\noindent where the $X_j[s]$, $s\in A^{r_i+1}$ are new  propositional variables (not used in $\calC$). 
If $\mA\models_{X_i}\exists x \varphi_j$ then, there exists a function $F\colon X_i\rightarrow \mathcal{P}(A)\setminus \{\emptyset\}$, such that $\mA\models_{X_i(F/x)} \varphi_j$. In other words,  $\mA\models_{X_j}\varphi_j$ for some team $X_j$ defined by the solutions  of the constraint $\bigwedge_{s\in A^{r_i}} X_i[s]\rightarrow \bigvee_{s'=(s,a),\  a\in A} X_j[s']$ (which define a suitable function $F$). Conversely, if $\mA\models_{X_j}\varphi_j$ for a team $X_j$ as above defined from $X_i$, then clearly $\mA\models_{X_i}\exists x \varphi_j$.

\textsf{Type of clauses:} all clauses that are introduced are dual-Horn (implication of the form 
$A\leftrightarrow B\lor C$ can be rewritten $(A\rightarrow B\lor C)\wedge(B\rightarrow A)\wedge(C\rightarrow A)$.

  	\item[($\forall$)] If $\varphi_i$ is   $\forall x \varphi_j$, then: $\calS := ( \calS \backslash \{(\varphi_i,X_i,r_i)\} ) \cup \{(\varphi_j,X_j, r_j=r_i+1)\} $ and 
  	  		\[
 \begin{array}{rl}
  \calC :=  \calC  \cup &  \{X_i[s]\leftrightarrow X_j[s']: \\
   & s\in A^{r_i}, s'\in A^{r_i+1} \mbox{ s.t. } \restrict{s'}{\tu x}=\restrict{s}{\tu x} \},  
 \end{array}
  	  	\]
%
%
%
  	  	
  	\noindent where the $X_j[s]$, $s\in A^{r_i+1}$ are new  propositional variables (not used in $\calC$). The conclusion is similar as for the preceding case.

	\textsf{Type of clauses:} All clauses that are introduced are both Horn, dual-Horn and $2$-CNF (they are said implicative $2$-CNF).	
  	
\end{description}

\noindent \textbf{($\wedge$)-rules.} If $\varphi_i$ is   $\varphi_j \wedge \varphi_h$ we may consider two main rules.
  	
  	\begin{enumerate}[label=$\wedge.\arabic*$,ref=($\wedge$.\arabic*)]
  	    \item~\label{rule: wedge} If $\varphi_i$ is   $\varphi_j \wedge \varphi_h$ then: $X_i=X_j=X_h$, $r_i=r_j=r_h$,  $\calS := ( \calS \backslash \{(\varphi_i,X_i,r_i)\} ) \cup \{(\varphi_j,X_j, r_j), (\varphi_h,X_h, r_h)\}$
  	and $\calC$ just states the equality between all teams involved. Then, 	 
	  $$
	  \begin{array}{lr}
	  \calC :=  \calC & \cup \{X_i[s]\leftrightarrow X_j[s]: s\in A^{r_i} \}\\
	  				& \cup \{X_i[s]\leftrightarrow X_h[s]: s\in A^{r_i}\} 
	  \end{array}
	  $$

  	By definition, $\mA\models_{X_i}\varphi_i$ iff $\mA\models_{X_i}\varphi_j \wedge \varphi_h$.

	  \textsf{Type of clauses:} The clauses that are introduced are all Horn, Dual-Horn and 2-CNF. They are not $\oneequivDH$.
	  
	 \item~\label{rule: wedge FO}  If $\varphi_i$ is   $\varphi_j \wedge \varphi_h$ with $\varphi_h\in \fo$ then $X_j$ can be computed in $\logspace$, $r_i=r_j=r_h$ and $\calS := ( \calS \backslash \{(\varphi_i,X_i,r_i)\} ) \cup \{ (\varphi_h,X_h, r_h)\}$
  	and $\calC$ just states the equality between all teams involved. Then, 	 $$\textstyle
		   \calC :=  \calC \cup 
   \{X_i[s]\leftrightarrow X_j[s]: s\in A^{r_i}\cap X^j \} 
					 $$
  	
	  \textsf{Type of clauses:} The clauses that are introduced are all Horn, Dual-Horn and 2-CNF. They also are $\oneequivDH$.
	  
  	\end{enumerate}
  	
  Note that these rules preserve $\logspace$ computability of the teams: if $X_i$ is $\logspace$ computable then so are $X_h=X_j$ since there are equal to $X_i$.

\noindent\textbf{($\vee$)-rules} 	If $\varphi_i$ is   $\varphi_j \vee \varphi_h$, we distinguish one general case and two particular ones.
	   
	\begin{enumerate}[label=$\vee.\arabic*$,ref=($\vee$.\arabic*)]
	\item\label{rule:general or} In the general case, we set\footnote{Mainly for convenience we do not adjust $r_j$ and $r_j$ to the real number of free variables of $\varphi_j$ and $\varphi_h$ and keep conserving the majoration by $r_i$} $r_i=r_j=r_h$ and  $\calS := ( \calS \backslash \{(\varphi_i,X_i,r_i)\} ) \cup \{(\varphi_j,X_j, r_j), (\varphi_h,X_h, r_h)\}$ and
		$$
		\begin{array}{rl}
		\calC :=  \calC &\cup 
   \{X_i[s]\rightarrow X_j[s]\vee X_h[s]: s\in A^{r_i} \}\\
   						& \cup \{X_j[s]\rightarrow X_i[s],  X_h[s]\rightarrow X_i[s]: s\in A^{r_i} \} 
		\end{array}
		$$
		
	 \noindent where again the $X_j[s]$ and $X_h[s]$, $s\in A^{r_i}$ are  new propositional variables  (not used in $\calC$). Here again, $\mA\models_{X_i}\varphi_i$ if and only if $\mA\models_{X_j}\varphi_j$ and $\mA\models_{X_h}\varphi_h$ for some suitable $X_j$ and $X_h$ such that $X_j\cup X_h=X_i$ which is exactly what is stated by the Boolean constraints.
  
	 \textsf{Type of clauses:} the first set of clauses that are introduced is made of dual-Horn clauses. The second set if made of clauses which are both Horn, dual-Horn and $2$-CNF.

	 \item\label{rule:vee general+fo} If 
	 $\varphi_j\in\fo$. Let $X_j=\{s:  (\mA,s')\models \varphi_j \mbox{ with } s'=\restrict{s}{\free{\varphi_j}}\}$. Since $\varphi_j\in \fo$, the set $X_j$ can be computed in $\logspace$. One then set: $r_i=r_j=r_h$, $\calS := ( \calS \backslash \{(\varphi_i,X_i,r_i)\} ) \cup \{(\varphi_h,X_h, r_h)\} $ and
	 	$$
		\begin{array}{rl}
	   \calC :=  \calC & \cup \{X_i[s]\rightarrow  X_h[s]: s\in A^{r_i}\backslash X_j \}\\
					& \cup \{ X_h[s]\rightarrow X_i[s]: s\in A^{r_i} \}$$
		\end{array}
		$$
 \textsf{Type of clauses:} in this case, the clauses that are introduced are both Horn, dual-Horn and $2$-CNF.

	\item\label{rule:vee general+logspace} If $X_i$ is logspace computable from $X$ and $\mA$, then we set $r_i=r_j=r_h$ and $\calS := ( \calS \backslash \{(\varphi_i,X_i,r_i)\} ) \cup \{(\varphi_j,X_j, r_j), (\varphi_h,X_h, r_h)\} $ and
	$$\textstyle
 \calC :=  \calC \cup 
\{X_j[s]\vee X_h[s]: s\in X_i \}
$$

\textsf{Type of clauses:} in this case, the clauses that are introduced are both dual-Horn and $2$-CNF.

\item\label{rule:vee logspace+fo+downward} If 
$\varphi_j\in \fo$, $X_i$ is $\logspace$ computable and $\varphi_h$ is downward closed. We proceed as in the preceding case just noting that, since $\varphi_h$ is downward closed we can set $X_h=X_i\backslash X_j$.  One then set: $r_i=r_j=r_h$, $\calS := ( \calS \backslash \{(\varphi_i,X_i,r_i)\} ) \cup \{(\varphi_h,X_h, r_h)\} $ and
	 $$\textstyle
  \calC :=  \calC \cup 
\{ X_h[s]: s\in X_i\backslash X_j \}
$$

\textsf{Type of clauses:} in this case, the clauses that are introduced are all unitary (hence obviously both Horn, dual-Horn and $2$-CNF).

Note that in this case, $X_h$ is also $\logspace$ computable.

\end{enumerate}

\noindent\textbf{($\vee_x^b$)-rules.}  We examine here the cases where  $\varphi_i=\varphi_j \vee_x \varphi_h$ or $\varphi_i=\varphi_j \vee^b_x \varphi_h$.

\begin{enumerate}[label=$\vee^b_x.\arabic*$,ref=($\vee^b_x$.\arabic*)]
    
    \item\label{rule: or relativized block general} Let $\varphi_i=\varphi_j \vee^b_x \varphi_h$. We set $r_i=r_j=r_h$ and  $\calS := ( \calS \backslash \{(\varphi_i,X_i,r_i)\} ) \cup \{(\varphi_j,X_j, r_j), (\varphi_h,X_h, r_h)\}$ and
			
			\[
		   \calC :=  
		   \begin{array}[t]{l}
	\calC \cup 
   \{X_i[s]\leftrightarrow X_j[s]\vee X_h[s]: s\in A^{r_i} \}\cup \\ 
   \{X_j[s]\leftrightarrow X_j[s'] : s,s'\in A^{r_j}, s(x)=s'(x) \} \cup \\
   \{X_h[s]\leftrightarrow X_h[s'] : s,s'\in A^{r_h}, s(x)=s'(x) \}
    \end{array}\]
		
	 \noindent where the $X_j[s]$ and $X_h[s]$, $s\in A^{r_i}$ are  new propositional variables. 
  
	 \textsf{Type of clauses:} the set of clauses that are introduced are  dual-Horn clauses. 
  
    \item\label{rule: or relativized block logspace}  Let $\varphi_i=\varphi_j \vee^b_x \varphi_h$ and $X$ be computable in $\logspace$. We set $r_i=r_j=r_h$ and  $\calS := ( \calS \backslash \{(\varphi_i,X_i,r_i)\} ) \cup \{(\varphi_j,X_j, r_j), (\varphi_h,X_h, r_h)\}$ and
			
			\[
		   \calC :=  
		   \begin{array}[t]{l}
	\calC \cup 
   \{X_j[s]\vee X_h[s]: s\in X_i \}\cup \\ 
   \{X_j[s]\leftrightarrow X_j[s'] : s,s'\in A^{r_j}, s(x)=s'(x) \} \cup \\
   \{X_h[s]\leftrightarrow X_h[s'] : s,s'\in A^{r_h}, s(x)=s'(x) \}
    \end{array}\]
		
	 \noindent where the $X_j[s]$ and $X_h[s]$, $s\in A^{r_i}$ are  new propositional variables. 
  
	 \textsf{Type of clauses:} the set of clauses that are introduced are both $2$-CNF and dual-Horn clauses. 
\end{enumerate}

  \medskip

\noindent\textbf{Operator rules.}

  \begin{description}
	  \item[(inc)-rule] If $\varphi_i$ is  $\tu x \subseteq \tu y$ then: $\calS:= \calS\backslash \{(\varphi_i,X_i,r_i)\}$  and  
  	\[
  		\calC:=\calC \cup \{X_i[s]\rightarrow \bigvee_{s'\in A^r, \restrict{s'}{\tu y}=\restrict{s}{\tu x}} X_i[s']:
  	 s\in A^r \}. \]

  	It holds that $\mA\models_{X_i}\tu x \subseteq \tu y$ iff $ \bigwedge_{s\in A^{r_i}} (X_i[s]\rightarrow \bigvee_{s'\in A^{r_i}, \restrict{s'}{\tu y}=\restrict{s}{\tu x}} X_i[s'])$ is satisfiable.

	  \textsf{Type of clauses:} all the clauses that are introduced are dual-Horn.

	  \item[(anon)-rule] If  $\varphi_i$ is $\anon{\tu x}{\tu y}$ then: $\calS:= \calS\backslash \{(\varphi_i,X_i,r_i)\}$  and  
  	\[
  		\calC:=\calC \cup \{X_i[s]\rightarrow \bigvee_{\substack{s'\in A^{r_i},\\ \restrict{s'}{\tu x}=\restrict{s}{\tu x} \wedge \restrict{s'}{\tu y}\neq \restrict{s}{\tu y}}} X_i[s']:
  	 s\in A^r \}. \]

  	It holds that $\mA\models_{X_i}\anon{\tu x}{\tu y}$ iff $ \bigwedge_{s\in A^{r_i}} (X_i[s]\rightarrow \bigvee_{s'\in A^{r_i}, \restrict{s'}{\tu x}=\restrict{s}{\tu x} \wedge \restrict{s'}{\tu y}\neq \restrict{s}{\tu y}} X_i[s']$ is satisfiable.

	  \textsf{Type of clauses:} all the clauses that are introduced are dual-Horn.

	  \item[(dep)-rule] If $\varphi_i$ is  $\dep(\tu x,y)$ then: $\calS:= \calS\backslash \{(\varphi_i,X_i,r_i)\}$  and  
  	\[
  		\calC:=\calC \cup \{\bigwedge_{ \substack{s,s'\in A^{r_i}\\ s(\tu x)=s'(\tu x) \wedge s(y)\neq s'(y)}} (\neg X_i[s]\vee \neg X_i[s']) \}. \]

		  \textsf{Type of clauses:} all the clauses that are introduced are both Horn and $2$-CNF.
  

		\item[(indep)-rule] If $\varphi_i$ is $\indep{\tu x}{\tu y}{\tu z}$ then: $\calS:= \calS\backslash \{(\varphi_i,X_i,r_i)\}$  and  
	\[
	\begin{array}{lr}\displaystyle
	\calC:=\calC \cup & \{\bigwedge_{\tu a\in A^{|\tu y|}}	\bigwedge_{\substack{s,s'\in A^{r_i}\\ s(\tu y)=s'(\tu y)=\tu a}} ( X_i[s] \wedge X_i[s'] \rightarrow  \\
	  & \bigvee_{\substack{s'' \in A^{r_i},  s''(\tu y)=\tu a \\ s''(\tu x)=s(\tu x)\wedge  s''(\tu z)=s'(\tu z)}} X_i[s''])  \}. \end{array} \] 
		
	\textsf{Type of clauses:} the clauses are not of any identifiable cases

\item[(excl)-rule] If $\varphi_i$ is $\tu x|\tu y$ then: $\calS:= \calS\backslash \{(\varphi_i,X_i,r_i)\}$  and

\[
		\calC:=\calC \cup \{\bigwedge_{ \substack{s,s'\in A^{r_i}:  s(\tu x)=s'(\tu y)}} (\neg X_i[s]\vee \neg X_i[s']) \}. \]

		\textsf{Type of clauses:} all the clauses that are introduced are both Horn and $2$-CNF.

\item[(const)-rule] The constancy operator $\dep{(x)}$ is a particular case of dependence and, as such, can be translated by the same set of clauses. However, a specific treatment can be done that will not ends up by the addition of new conjunctive clauses but by the addition of a "small" size formula in disjunctive normal form. So, If $\varphi_i$ is  $\dep(x)$ then: $\calS:= \calS\backslash \{(\varphi_i,X_i,r_i)\}$  and  
  	\[
  		\calC:=\calC \cup \{\bigvee_{a\in A} \bigwedge_{b\in A\backslash\{a\}} \bigwedge_{\substack{s\in A^{r_i}\\ s(x)=b}} \neg X_i[s]\}. \]

\end{description}

  Observe that applied to some $(\varphi_i,X_i,r_i)$, the algorithm above only adds triples  $\calS$ whose first component is a proper subformula of $\varphi_i$ and eliminates  $(\varphi_i,X_i,r_i)$. When the formula $\varphi_i$ is atomic, no new triple is added afterwards.  Hence the algorithm will eventually terminate with $\calS=\emptyset$. 
 Setting $\Psi:= \exists \overline{X_1[\cdot]}\ldots \exists \overline{X_t[\cdot]}\bigwedge_{C\in\calC} C$, with each $\overline{X_i[\cdot]}$ ranges over propositional variables $X_i[s]$ for $s\in A^{r_i}$, it can easily be proved by induction that: $\mA \models_X \varphi$  iff  $\Psi$ is satisfiable.


Observe also that each clause in $\calC$ can be constructed from $X$ and $\mA$ by simply running through their elements (using their index) hence  in logarithmic space. 

 \section{Deriving tractable fragments from $\textsc{SAT}$ translation}~\label{sec:Deriving tractable fragments}

 This section relies on the methods of Theorem~\ref{theorem: translation} to exhibit (or prove alternatively) new tractability results for fragments of team-based logics.

\subsection{Anonymity, constancy and inclusion logic}

The complexity of  $\FO{\subseteq}$ is well understood:  $\FO{\subseteq}$-sentences define $\Ptime$-properties and the logic captures $\Ptime$ on finite ordered structures \cite{gallhella13}. In this section, we show that the class can be substantially extended while preserving inclusion in $\Ptime$, in particular anonymity and constancy atoms as well as block disjunction can be added.  Let $\FO{\vee_x^b,\subseteq, \anon{}{},\const}$ be defined by the following grammar, where $\alpha\in \fo$:

\[
\varphi:=\alpha \ |\  \tu x\subseteq \tu y \ |\  \anon{\tu x}{\tu y} \ | \dep{(x)}\ |\  \varphi\vee\varphi \ |\  \varphi\vee^b_x\varphi \ |\  \varphi\wedge\varphi \ |\  \exists x\varphi \ |\  \forall x\varphi,
\]

\begin{proposition}~\label{prop: fo incl anon const to union o dual horn}
Let $\varphi\in\FO{\vee_x^b,\subseteq, \anon{}{},\const}$. There is a logspace algorithm which on input $\mA$ and $r\in \mathbb{N}$ with $r\geq |\Fr{(\varphi})|$, outputs a propositional formula $\Psi$ which is a disjunction of dual Horn formulas 
with $\mX=\Fr{(\Psi)}\subseteq \var{\Psi}$ such that, for all teams $X\subseteq A^r$, the following are equivalent:
\begin{itemize}
    \item $\mA\models_X\varphi$
    \item There is a assignment $I:\mX\longrightarrow \{0,1\}$ that satisfies $\Psi$ such that for all $X[s]\in \mX$:  $I(X[s])=1 \iff s\in X$.
\end{itemize}
If $\varphi\in\FO{\vee_x^b,\subseteq, \anon{}{}}$, then the output $\Psi$ is dual-Horn.
\end{proposition}

\longversion{
\begin{proof}
Considering the translation to $\pb{SAT}$ starting from a formula $\varphi$ in $\FO{\vee_x,\subseteq, \anon{}{},\const}$, one only need to use the following rules:  ($\fo$),($\exists$), ($\forall$),\ref{rule: wedge}, \ref{rule:general or}, ($\vee_x^b$), (inc), (anon) and (const). The formula $\Psi$ obtained after the translation verifies that: $\mA\models_X\varphi$ iff $\Psi$ is satisfiable. Let $k$ be the number of constancy atoms in $\varphi$, formula $\Psi$ is of the form:
    
    \[
    \Psi\equiv\exists \tu p \bigwedge_{i=1}^n C_i \wedge \Phi_1 \wedge \cdots \wedge\Phi_k
    \]
    
    \noindent where $n$ depends on the size of $\mA$ and $X$, each $C_i$ is dual-Horn and each $\Phi_j$, $j=1,...,k$, is a DNF formula $\bigvee_{a_j\in A}F_j$, with $F_{j,a_k}$ being a polynomial size conjunction of literals. Hence $\Psi$ can be rewritten as:
    
    \[
    \Psi\equiv\bigvee_{a_1\in A}\cdots\bigvee_{a_k\in A} \exists \tu p\bigwedge_{i=1}^n C_i \wedge F_{1,a_1}\wedge \cdots \wedge F_{k,a_k}.
    \]
    
    \noindent i.e. as a polynomial size number of disjunctions of dual-Horn formulas. 
\end{proof}
} 

The above result has consequences on the complexity of many reasoning tasks for this logic.

 \begin{corollary}~\label{prop: fo incl anon} 
\begin{enumerate}
    \item 
    $\FO{\vee_x^b,\subseteq, \anon{}{},\const}\subseteq \Ptime$.
    \item $\cTeam{\FO{\vee_x^b,\subseteq, \anon{}{},\const}}\subseteq \cP$.
    \item For any  $\varphi\in \FO{\vee_x^b,\subseteq, \anon{}{},\const}$, there is an algorithm which upon input $\mA$, enumerates the teams $X\subseteq A^{|\free{\varphi}|}$ satisfying $\varphi$ with polynomial delay and polynomial space.
    \item Let $\varphi\in\FO{\vee_x^b,\subseteq, \anon{}{},\const}$, Given $\mA$ and $X$, one can compute all maximal subteams $X'\subseteq X$ such that $\mA\models_{X'}\varphi$ in polynomial time.
\end{enumerate}
\end{corollary}
In the subsequent sections we state our results  only for model-checking but, analogously to Corollary \ref{prop: fo incl anon},  results for the complexity of counting and enumeration also follow.
\longversion{
\begin{proof}
\begin{enumerate}
    \item The team $X$ is part of the input of the mode-checking problem, so we have to add the constraint from rule \ref{rule:init} to the formula obtained in Proposition~\ref{prop: fo incl anon const to union o dual horn}. It is know that  deciding the satisfiability of dual Horn formulas can be done in polynomial time. If $\Psi=\bigvee_{i\in I}\Psi_i$ where each $\Psi_i$ is dual Horn then one can test the satisfiability of $\Psi$ by running successively the polynomial time algorithm for the $\Psi_i$ until one is found satisfied or all are found unsatisfiable. Completeness is obtained for already very simple formulas in $\FO{\subseteq}$ (see~\cite{gallhella13}).
    \item  Let $\varphi \in \FO{\vee_x^b,\subseteq, \anon{}{},\const}$ and   denote by $\mX_{(\mA,X)}$ and $\Psi_{(\mA,X)}$ the outputs of the logspace algorithm of \ref{prop: fo incl anon const to union o dual horn} with input $(\mA,X)$. Define the language 
    $$ L=\{(I,\Psi)\ | I\colon \mX_{(\mA,X)}\rightarrow \{0,1\}, \ I\models\Psi, \ \exists p I(p)=1 \}.$$  
    Now clearly $L$ is polynomially balanced, gives rise to the same function as $\varphi$, and witnesses containment in $\cP$.
    \item To the formula obtained in Proposition~\ref{prop: fo incl anon const to union o dual horn} add \ref{rule:init union} (instead of \ref{rule:init} above). In this constraint set, one expresses that if a tuple $s$ is not in $X$ then $X[s]$ must be false. The status of propositional variables $X[s]$ with $s\in X$ is left open. Thus, there is a bijection between the subteams $X'$ of $X$ such that $\mA\models_{X'} \varphi$ and the maximal of propositional variables $X[s], X_1[s], ..., X_t[s]$, fixing $X[s]=1$ if $s\in X'$, $X[s]=0$ if $s\in X\backslash X'$ that satisfy $\Psi$. One can then use the fact that given a union of dual Horn formulas $\exists \tu Y\Psi(\tu X, \tu Y)$, enumerating the distinct assignments $s:\tu X\longrightarrow\{0,1\}$ for variables in $\tu X$ such that $\Psi(s(\tu X), \tu Y)$ is satisfiable can be done in polynomial delay and memory space. Indeed, if $\Psi$ is a disjunction of dual Horn formulas and $v\in \Fr{(\Psi)}$, then formulas $\Psi(0/v)$ and $\Psi(1/v)$ are also disjunctions of dual Horn formulas. Hence, one can then develop a recursive algo that assigns values step by step  to the $\tu X$ variables only and check for satisfiability after each such variable assignment to determine if a satisfying assignment can still be found and cut immediately in the search tree if not. (see~\cite{CreignouH97} for the particular case of enumerating satisfying assignments of dual Horn formulas). 
    
    \item Suppose first that, $\varphi\in \FO{\vee_x^b,\subseteq, \anon{}{}}$. In the translation, again just  just use \ref{rule:init union} instead of \ref{rule:init}. We end up with a dual Horn propositional formula. In this constraint set, one expresses that if a tuple $s$ is not in $X$ then $X[s]$ must be false. The status of propositional variables $X[s]$ with $s\in X$ is left open.  It is well known that computing the (unique) maximal model of a propositional formula in dual Horn form $\Psi$ can be done in polynomial time: First assign the unitary clauses (with $true$ or $false$ depending on whether they are positive or negative and continue to do so as long as unitary clauses appear after substitution. If a contradiction does not emerge then the formula is made of dual horn clauses, none of which is unitary. Then, completing the assignment by mapping each remaining variable to $true$ yields a satisfying and maximal assignment). In this case, it is also maximal in the variable $X[s]$ only.
    The case of $\varphi\in\FO{\vee_x^b,\subseteq, \anon{}{},\const}$ is similar as above instead that the translation ends up with a disjunction of dual-Horn formulas. Such formulas may not have one single maximal models only but a polynomial number. One just have to try in turn each dual Horn subformulas and compute their maximal models.
\end{enumerate}	
\end{proof}
} 

\subsection{Tractable fragment of anonymity, constancy and inclusion logic}~\label{section: weak fragment inclusion logic}

Let $\FO{\subseteq, \anon{}{},\const}_w$ be the fragment of $\FO{\subseteq, \anon{}{},\const}$ defined by the following grammar:

\[
\varphi:=\alpha \ |\  \tu x\subseteq \tu y \ |\  \anon{\tu x}{\tu y} \ | \ \dep(x)\ |\  \varphi\vee\varphi  \ |\  \varphi\wedge\alpha \ |\  \exists x\varphi 
\]

\noindent where $\alpha\in \fo$. The fragment for inclusion atom only, $\FO{\subseteq}_w$, has been defined in~\cite{HannulaH19} where it is proved it has a model-checking 
in $\nl$. We prove here that the result extends to  $\FO{\subseteq, \anon{}{},\const}_w$.

\begin{proposition}
$\FO{\subseteq, \anon{}{},\const}_w$ is in $\nl$.
\end{proposition}

\longversion{
\begin{proof}
In the translation to $\pb{SAT}$ starting from a formula $\varphi\in \FO{\subseteq, \anon{}{},\const}_w$, one only need to use the following rules from the framework develop for proving Theorem~\ref{theorem: translation}: \ref{rule:init}, ($\fo$),($\exists$), \ref{rule:general or}, \ref{rule: wedge FO}, (inc), (anon) and (const). One obtains a propositional formula $\Psi$ of the form:

    \[
    \Psi\equiv\exists \tu p\bigwedge_{i=1}^n C_i \wedge \Phi_1 \wedge \cdots \wedge \Phi_k
    \]
    
    \noindent where $k$ is the number of constancy atoms in $\varphi$, $n$ depends on the size of $\mA$ and $X$ and such that $\bigwedge_{i=1}^n C_i$ is of $\oneequivDH$ form and each $\Phi_j$, $j=1,...,k$, is a DNF formula $\bigvee_{a_j\in A}F_j$, with $F_{j,a_k}$ being a polynomial size conjunction of literals. Hence $\Psi$ can be rewritten, in logarithmic space, as:
    
    \[
    \Psi\equiv\bigvee_{a_1\in A}\cdots\bigvee_{a_k\in A} \exists \tu p\bigwedge_{i=1}^n C_i \wedge F_{1,a_1}\wedge \cdots \wedge F_{k,a_k}.
    \]
    
    \noindent Since each $F_{j,a_j}$ is a conjunction of unit clauses, $\Psi$ is a disjunction of  $\oneequivDH$ formulas. Thus, $\Psi$ is satisfiable iff one can find $a_1,\ldots, a_k$ in $A$ such that $\bigwedge_{i=1}^n C_i \wedge F_{1,a_1}\wedge \cdots F_{k,a_k}$ is satisfiable. Each inner test is in $\nl$ by Proposition~\ref{prop: EDH in NL} and keeping tr of the current value of $a_1,...,a_k$ can be done with logarithmic space. Hence, the whole satisfiability test is in $\nl$. 
\end{proof}
} 

Note that, from~\cite{HannulaH19}, $\nl$-completeness is already obtained for the model-checking of the very simple formula of $\FO{\subseteq}_w$ consisting in the disjunction of two inclusion atoms with disjoint variables $x\subseteq y \vee u\subseteq v$ and even for formula $x\subseteq y \vee u=v$. 
On the other side, it is also known that for formulas  $(x\subseteq z \wedge y\subseteq z) \vee u\subseteq v$ and $(x\subseteq z \wedge y\subseteq z) \vee u = v$, the model-checking problem is $\Ptime$-complete. 
Hence, together, with the fragment made of conjunctions of inclusion and anonymity atoms only (which is in $\fo$ hence in $\logspace$),  $\FO{\subseteq, \anon{}{},\const}_w$ is one of the maximal fragment whose model-checking is in $\nl$.

\subsection{Tractable fragments of dependence and exclusion logic}~\label{section: weak fragment dependence logic}
It was shown in \cite{10.1145/3471618} that dependence logic without $\vee$ and $\exists$ collapses to $\nl$. We can extend this result using our methods to allow a weak use of disjunction. Define
 $\FO{\dep(),|}_w$ by : 
\[\varphi:=\alpha \ |\  \dep(\tu x,y) \ \ |\  \tu x|\tu y \ | \ \varphi\wedge\varphi  \ |\  \varphi\vee\alpha \ |\  \forall x\varphi,\]
\noindent where $\alpha\in \fo$.

 \begin{proposition}\label{prop: weak fragment dep}
$\FO{\dep(\cdot,\cdot),|}_w\subseteq \nl$.
\end{proposition}

\begin{proof}
 From the translation, all clauses that are generated are $2$-CNF by the use of rules: \ref{rule:init}, ($\fo$), ($\forall$), \ref{rule:vee general+fo}, \ref{rule: wedge}, (dep), (excl).
\end{proof}

The above tractability result can be further extended via the notion of  disjunction width (defined in \cite{10.1145/3471618}). 

\begin{definition} Let $\varphi\in \FO{\dep(\cdot,\cdot),|}$, the disjunction width of $\varphi$, denoted $\Disjunctiondepth{\varphi}$, is defined as follows:
	\[
	\Disjunctiondepth{\varphi}=
	\left\{\begin{array}{l}
	1  \mbox{ if } \varphi  \mbox{ is }   \dep(\tu x, y) \mbox{ or } \tu x | \tu y \\
	0   \mbox{ if } \varphi \in \fo \\
	\max (\Disjunctiondepth{\varphi_1}, \Disjunctiondepth{\varphi_2}) \mbox{ if } \varphi \mbox{ is } \varphi_1\wedge\varphi_2 \\
	\Disjunctiondepth{\varphi_1} + \Disjunctiondepth{\varphi_2} \mbox{ if } \varphi \mbox{ is } \varphi_1\vee\varphi_2 \\
	\Disjunctiondepth{\varphi_1}  \mbox{ if } \varphi \mbox{ is }\forall x \varphi_1\\
	 \infty  \mbox{ if } \varphi \mbox{ is } \exists x \varphi_1
	\end{array}\right.
	\]
	\end{definition}

Remark that if $\varphi\in \FO{\dep(\cdot,\cdot),|}_w$, then $\Disjunctiondepth{\varphi}\leq 1$.
\longversion{The following  formula is of disjunctive width bounded by $2$:

\[\forall x_1,..., x_n (\bigwedge_{i=1}^n \dep(\tu x_i,y_i)\vee\bigwedge_{i=1}^m  z_i | t_i)\vee \varphi ) \wedge \psi\]

\noindent where $\varphi,\psi\in\fo$.
} 
Let $\mathcal{L}_{w,2}$ consist of  $\varphi\in \FO{\dep(\cdot,\cdot),|}$ with $\Disjunctiondepth{\varphi}\le 2$. 
\begin{proposition}\label{p20}
$\mathcal{L}_{w,2}\subseteq \nl$.
\end{proposition}
A weaker version of this result for quantifier-free dependence logic formulas was shown in \cite{10.1145/3471618}.
\longversion{
\begin{proof}
Let $\mA$ a structure, $\varphi$ a formula as above and $X$ a team. Let $T(\varphi)$ be the syntactic tree of $\varphi$.  Note that $\varphi$ is downward closed. By hypothesis, on every paths from a leaf to the root of $T(\varphi)$, there  can be at most one node $i$ representing a subformula $\varphi_i$ of the form $\varphi_i=\varphi_j\vee \varphi_h$ with $\Disjunctiondepth{\varphi_j}=\Disjunctiondepth{\varphi_h}=1$. W.l.o.g. let us call $\varphi=\varphi_1,...,\varphi_i$ the subformulas from the root to $\varphi_i$.  By definition, for $a=1,...,i-1$, the root of subtree $T(\varphi_a)$ is either labeled by :

\begin{itemize}
    \item a disjunction  $\varphi_{i_{a+1}}\vee \varphi_a'$ with $\varphi_{i_{a+1}}$ being downward closed, $\Disjunctiondepth{\varphi_{a+1}}=2$ and $\Disjunctiondepth{\varphi_a'}=0$ i.e. $\varphi_a'\in \fo$
    \item a conjunction $\varphi_{i_{a+1}}\wedge \varphi_a'$
    \item a universal quantification $\forall x \varphi_{i_{a+1}}$
\end{itemize}

In a translation to $\pb{SAT}$, one can then use rules for~\ref{rule:vee logspace+fo+downward}, conjunction and universal quantification.
All these cases preserve $\logspace$ computability of accompanying teams so $X_i$ must be $\logspace$ computable. At this point, rule \ref{rule:vee general+logspace} can be used. To treat $\varphi_j$ and $\varphi_h$ subtrees, one remark that, from now, all disjunctions involve again at least one $\fo$ formula and that other subformulas that appear uses conjunction, universal quantification, first-order formulas, dependence or exclusion atoms. Putting all of this together we obtain a translation into a propositional formula in $2$-CNF.  
\end{proof}
} 



\subsection{The case of independence logic}

The next result shows that the analogues of Propositions \ref{prop: weak fragment dep} and \ref{p20} and not true for independence logic. 

\begin{proposition}\label{prop: weak fragment indep}
The model-checking of the following formula of disjunction width 1 is already $\np$-complete: 

\[
w=0 \vee (c_1\perp_c c_2 \wedge x \perp_z x).
\]
\end{proposition}

\longversion{
\begin{proof}
	Let $\varphi_1\equiv w=0$, $\varphi_2\equiv c_1\perp_c c_2 \wedge x \perp_z x $. We will reduce  $3$-SAT to the  model-checking problem  of $\varphi=\varphi_1\vee\varphi_2$. 
	Let $\Phi = \bigwedge_{i=1}^n C_i$ be  a $3$-SAT instance on the set of variables $\{v_1,...,v_m\}$. Each $C_i= p_{i_1} \vee p_{i_2} \vee p_{i_3}$  is a disjunction of $3$ literals
	where  each $p_{i_j}\in \{\, v_{i_j},\neg v_{i_j}\}$. To this instance we associate a universe $\mA$ and a team $X$ on the variables $w,c,c_1,c_2,z,x$. The universe $\mA$ is  $\{\, v_1,\hdots, v_m,\neg v_1,\hdots, \neg v_m\,\}\cup\{\,0,...,n\,\}$. For each clause $C_i$ we add  in $X$ the $6$ assignments displayed on the below:
	\[
	\begin{array}{|c|c|c|c|c|c|}
	\hline
	w&c&c_1&c_2&z&x\\
	\hline \hline
	0&i&1&1&v_{i_1}&p_{i_1}\\
	0&i&1&1&v_{i_2}&p_{i_2}\\
	0&i&1&1&v_{i_3}&p_{i_3}\\
	\hline
	1&i&0&0&0&0\\
	1&i&1&0&0&0\\
	1&i&0&1&0&0\\
	\hline
	\end{array}
\]

	We will next show that $\Phi$ is satisfiable if and only if $\mA \models_X \phi$.

	\begin{itemize}
		\item[$\Rightarrow$] Suppose there is an assignment  $I:\{\,v_1,\hdots,v_m\,\} \rightarrow \{0,1\}$ that evaluates  $\Phi$ to true, i.e., one literal in each clause (at least) is evaluated to $1$. We have to split $X$ into two sub-teams $X = Y \cup Z$ such that $\mA \models_Y w=0$ and $\mA \models_Z \left(c_1\perp_c c_2 \wedge x \perp_z x\right)$. We must put every clause-assignment $s \in X$ such that $s(w) = 1$ in $Z$. There are exactly three such assignments per clause..
		Since $I$ is a satisfiable assignment, there is at least one team assignment $s$ in each block such that  $s(w)=0$ and $s(x) = v_i$ if $I(v_i) = 1$, and $s(x) = \neg v_i$ if $I(v_i) = 0$. Note in passing that this assignment satisfy $s(c,c_1,c_2) = (i,1,1)$.  We  put such assignment in $Z$ and the other two  in $Y$. 
		
		It is easily seen that, such choices still implies that  $\mA \models_Y w=0$. 
		
		Since $I$ is a valuation, we clearly have $\mA \models_Z x \perp_z x$. It is easily seen that for each value $i$ of $c$, we now have  four assignments whose values for $(c,c_1,c_2)$ are respectively:  $(i,1,1), (i,1,0),(i,0,1)$ and $(i,0,0)$. Thus $\mA \models_Z c_1\perp_c c_2$.

		\item[$\Leftarrow$] Suppose then that there is a splitting  of $X$ into $Y,Z$ such that $X = Y \cup Z$, $\mA \models_Y w=0 $ and $\mA \models_Z \left(c_1\perp_c c_2 \wedge x \perp_z x\right)$.  Since $\mA \models_Y w=0 $, the assignments $(1,i,0,0,0,0)$, $(1,i,1,0,0,0)$ and $(1,i,0,1,0,0)$ are all in $Z$ and since $\mA \models_Z c_1\perp_c c_2$, for each clause $i$, at least one assignment among $(0,i,1,1,v_{i_1},p_{i_1})$, $(0,i,1,1,v_{i_2},p_{i_2})$, $(0,i,1,1,v_{i_3},p_{i_3})$ is in $Z$. The choice of which assignments to put in $Z$ is guided 
		by the fact that one must have $\mA \models_Z  x \perp_z x$, that force, for each variable $v$, its value to be a constant i.e. the mapping to be an assignment of the variables. Hence,  a satisfying assignment for $\Phi$ can be defined.
	\end{itemize}
\end{proof}
} 



\subsection{Mixing all atoms}~\label{sec:mixing all atoms}
We now consider the complexity of  formulas  with potentially all of the above atoms.



\begin{proposition}\label{anon_dep_relativized}
Let $\calL$ be a logic such that:

\begin{itemize}
    \item $\calL$ is downward closed
    \item 
    $\calL\subseteq \Ptime$
\end{itemize}

\noindent Then, 
the model-checking problem for any fixed formula 
$\varphi\vee_x \psi$
with  $\varphi\in \FO{\vee_x^b,\subseteq, \anon{}{},\const}$ and $\psi\in \calL$ is in $\Ptime$.
\end{proposition}

\begin{proof} 
Let $\varphi\in \FO{\vee_x^b,\subseteq, \anon{}{},\const}$ and $\psi\in \calL$
Let $X$ be a team and $\mA$ be a structure. Let $\mX=\{X[s]: s\in X\}$ be a set of propositional variables.
 $\mA\models_{X} \varphi\vee_x \psi$ iff there exists $Y,Z\subseteq X$ such that $\mA\models_{Y} \varphi$ and $\mA\models_{Z} \psi$ and $Y(x) \cap Z(x) = \emptyset$.
 
 One can construct in logarithmic space a propositional formula $\Psi=\bigvee_{i\in I}\Psi_i$ a disjunction of dual Horn formulas $\Psi_i$ on variable set $\var{\Psi}$ with $\mX\subseteq \var{\Psi}$ and such that there is a bijection between:
 
\begin{itemize}
    \item the maximal subteams $X'\subseteq X$ such that $\mA\models_{X'}\varphi$
    \item the maximal assignments of variables in $\var{\Psi}$ with $X[s]=1$ if $s\in X'$, $X[s]=0$ if $s\in X\backslash X'$ that satisfy $\Psi$
\end{itemize}
 
 Let us now consider, each formula $\Psi_i\wedge C$ where $C$ is the following conjunction of dual Horn constraints:

\[\bigwedge_{ \substack{s,s'\in X\\ s(x)=s'(x)}} X[s]\implies X[s'].\]

Again, it is easy to check that  there is a bijection between:

\begin{itemize}
    \item the maximal subteams $X'\subseteq X$ such that $\mA\models_{X'}\varphi$ and such that $X'$ is an $x$-block
    \item the maximal assignments of variables in $\var{\Psi}$ with $X[s]=1$ if $s\in X'$, $X[s]=0$ if $s\in X\backslash X'$ that satisfy $\bigvee_{i\in I}\Psi_i\wedge C$
\end{itemize}

Since  $\bigvee_{i\in I}\Psi_i\wedge C$ is a disjunction of dual Horn formulas, one can then construct its list of maximal models $X_1,..., X_k$, $k\leq |I|$, in  time polynomial in $|\mA|$ and $|X|$.

 
 
  Hence, $\mA\models_{X} \varphi\vee_x \psi$ iff there exists an integer $i\leq k$ and a subteam $Z'$ such that $X\backslash X_i\subseteq Z'$ and $\mA\models_{Z'}  \psi$. But since $\calL$ is downward closed,  it suffices to check whether $\mA\models_{X\backslash X_i} \psi$. This can be done in $\Ptime$ too.
\end{proof}

\longversion{
The proposition above is true, and even simpler to prove, for formulas of the form $\varphi\vee \psi$ instead of $\varphi\vee_x \psi$.} 
A large number of database decomposition constraints can be expressed in the above settings (cf. Example~\ref{example: horizontal decomposition}). The following then holds\longversion{ from the result above and tractability result from the preceding sections}.

  \begin{corollary}
  	The 
  	model checking problem for any fixed 
  	$\varphi\vee_x^b \psi$  with $\varphi\in \FO{\vee_x^b,\subseteq, \anon{}{},\const}$ and $\psi\in \FO{\dep(\cdot,\cdot),|}$ such that $\Disjunctiondepth{\psi}\leq 2$  is in $\Ptime$.
  	\end{corollary}

\begin{definition} The Boolean closure of an independence atom by first-order formulas, 
	denoted $\bcindepfo$, is defined as follows:

	\begin{itemize} 
		\item Any independence atom $\indep{\tu x}{\tu y}{\tu z}$ is in $\bcindepfo$.
		
		\item If $\varphi\in \bcindepfo$, then for any formula $\phi\in \fo$, $\varphi \wedge \phi$ and $\varphi\vee \phi$ are in $\bcindepfo$. 
	\end{itemize}
\end{definition}

\longversion{
Let $\varphi\in \bcindepfo$. Up to permutation of disjunction and conjunction, $\varphi$ can be put into the following normalized form:

\[
\varphi\equiv ((\dots ((\indep {\tu x}{\tu z}{\tu y} \wedge \phi_1) \vee \psi_1)\wedge \dots ) \wedge \phi_k) \vee \psi_k
\]

Let $\mA$ be any structure, $\CCp=\bigcap_{i=1}^k \phi_i(\mA)$ and $\CCm=\bigcup_{i=1}^k \psi_i(\mA)$, where, $\phi_i(\mA)$
is the set of assignments $s\colon \Fr(\varphi)\rightarrow A$ such that
$\mA \models_s \phi_i$. We can restate the fundamental property for satisfiability of an independence atom in a team (and a structure) to tackle the case of $\bcindepfo$ formulas. It holds that, for any $\varphi\in\bcindepfo$, any team $X$ and structure $\mA$, $\mA\models_X\varphi$ if and only if: 

\begin{enumerate}
	
	\item[C1] for all $s\in X$: either $s\in [((..(\psi_1\wedge \phi_2) \vee \dots ) \wedge \phi_k) \vee \psi_k](\mA)$ or $s\in\CCp$, $s\not\in\CCm$ and

	\item[C2]  for all $s_1,s_2\in X$ such that $s_1,s_2\in \CCp$, $s_1,s_2\not\in \CCm$, and $s_1(\tu z)=s_2(\tu z)$,  there exists $s_3\in X$ such that: $s_3(\tu z)=s_1(\tu z), s_3(\tu x)=s_1(\tu x) \mbox{ and } s_3(\tu y)=s_2(\tu y)$. 
	
	
\end{enumerate}

The first item is true by exhaustive case distinction. The second one comes from the fact that if a tuple $s$ satisfies  $s\in\CCp$ and  $s\not\in\CCm$  then it is forced to be in the sub-team satisfying $\indep{\tu x}{\tu z}{\tu y}$.  

Two assignments $s_1,s_2$ as in the second item will be said \textit{compatible} for formula $\varphi$ and team $X$ and $s_3$ is called a \textit{witness} of $s_1,s_2$ (for formula $\varphi$).
} 

It has been proved in~\cite{kontinenj13} that the data complexity of $x_1\perp_{c_1} y_1 \vee x_2\perp_{c_2} y_2 \vee x_3\perp_{c_3} y_3$ is $\np$-complete and in~\cite{10.1145/3471618}, that, by contrast, that of   $\bcindepfo \vee \bcindepfo$ is in $\nl$. We show below that this latter fragment can be extended.

%

\begin{proposition}\label{anon_indep}
 The 
 model checking problem for formulas 
 $\varphi\vee \psi_1\vee \psi_2$  with $\varphi\in \FOanon$ and $\psi_1,\psi_2\in\bcindepfo$ is in $\Ptime$. If $\varphi\in \fo$, then 
 $\varphi\vee \psi_1\vee \psi_2$ is in $\nl$.\end{proposition}

\begin{proof}Assume $\varphi\in \FOanon$ and finite $\mA$ and $X$ are also given. The proof in both cases can be split into two phases of which the second one is identical. In the first phase we find a maximal subteam $X'\subseteq X$ of $X$ such that $\mA \models_{X'} \varphi$. For any $\varphi\in \FOanon$ this can be done in  $\Ptime$ and for  $\varphi\in \textrm{FO}$ already in $\nl$. The second phase is similar to the proof of 13 in \cite{10.1145/3471618} in which we reduce in log-space the question whether some $(X\setminus X') \subseteq U\subseteq X$ satisfies $\mA \models_{U} \psi_1\vee \psi_2$ to the $\nl$-complete problem $2$-\pb{sat}. The stated complexity upper bounds are then easily obtained from the worst case complexity of executing the two phases consecutively in both cases for $\varphi$.

Now for the  structure $\mA$, team $X$, and the maximal subteam $X'\subseteq X$ satisfying $\varphi$, we construct a $2$-\pb{cnf} propositional formula $\Phi$ such that 
\begin{equation}\label{indep-translation}
\exists U(X'\subseteq U \subseteq X)\big( \mA \models_U \phi_1\vee \phi_2\big) \iff \Phi \mbox{ is satisfiable.} 
\end{equation}

 Recall that if a team $V$ is such that  $\mA \models_V \phi_1\vee \phi_2$ then, there exists $Y, Z\subseteq V$ such that $Y\cup Z=V$ and $\mA \models_Y \phi_1\mbox{ and } \mA \models_Z \phi_2$.
Now for each assignment $s\in X$, we introduce two Boolean variables $Y[s]$ and $Z[s]$. Our Boolean formula $\Phi$ will be defined below with these  $2|X|$ variables the set of which is denoted by $\Var(\Phi)$. It will express that a subset $X'\subseteq U\subseteq X$  can be split into $Y$ and $Z$ in such a way that incompatible assignments do not appear in the same subteams. 
 
For each pair $s_i,s_j$ that are incompatible for $\phi_1$ on team $X$, one adds the $2$-clause: $\neg Y[s_i] \vee \neg Y[s_j]$. The conjunction of these clauses is denoted by $C_Y$.
Similarly, for each pair $s_i,s_j$ that are incompatible for $\phi_2$ on team $X$, one adds the clause: $\neg Z[s_i] \vee \neg Z[s_j]$ and call $C_Z$  the conjunction of these clauses.

Finally, the construction of $\phi$ is completed by adding the following conjunctions:

\[ C^0:= \bigwedge \{Y[s] \vee Z[s]:{s\in X'}\} \]
\[ C^1:= \bigwedge\{ \neg Y[s]:{s \textrm{ fails C1 for } \phi_1} \}   \]
\[ C^2:= \bigwedge\{ \neg Z[s]:{s \textrm{ fails C2 for } \phi_2}\}   \]

It is not hard to see  that the  formula
 $$\Phi\equiv \bigwedge _{0\le i \le 2}C^i\wedge C_Y\wedge C_Z$$
  can be constructed deterministically in log-space.
 It remains to show that the equivalence \eqref{indep-translation} holds.
 
Assume that the left-hand side of the equivalence holds. Then, there exists $X'\subseteq U \subseteq X$ and $Y,Z$ such that $Y\cup Z=U$, $\mA \models_Y \phi_1$ and  $\mA \models_Z \phi_2$.
%
 We construct a propositional assignment $I:\Var (\Phi)\rightarrow \{0,1\}$ as follows. For all $s\in Y$, we set $I(Y[s])=1$ and for all $s\in Z$, we set similarly $I(Z[s])=1$. It is now immediate that the all of the clauses in  $\bigwedge _{0\le i \le 2}C^i$ are satisfied by $I$ since $X'\subseteq Y\cup Z$.
 
 Let us consider a clause  $ \neg Y[s_i] \vee \neg Y[s_j]$ for an incompatible pair $s_i,s_j$. Then, $I(Y[s_i])=0$ or $I(Y[s_j])=0$ must hold. For a contradiction, suppose that $I(Y[s_i])=I(Y[s_j])=1$. Then since $\mA \models_Y \phi_1$ holds, by construction $s_i$ and $s_j$ must be compatible for $\phi_1$. Hence we get a contradiction and may conclude that $I$ satisfies $ \neg Y[s_i] \vee \neg Y[s_j]$. The situation is similar for each clause $\neg Z[s_i] \vee \neg Z[s_j]$. 
 
 For the converse, let us then assume that  $\Phi$ is satisfiable, and let $I:\Var (\Phi)\rightarrow \{0,1\}$ be a satisfying assignment for $\Phi$. Since $I\models C^0$, we get that $I(Y[s])=1$ or $I(Z[s])=1$ for all  $s\in X'$. Define 
   $$ X_{Y}=\{s : I(Y[s])=1\} \mbox{ and } X_{Z}=\{s : I(Z[s])=1\}.  $$
 Now $X'\subseteq X_Y\cup X_Z\subseteq X$ and clauses $C^1$  ($C^2$) ensure that each $s\in X_Y$ ($s\in X_Z$) satisfies condition C1 above. We have to extend  the sets $X_Y$ and $X_Z$ to sets $Y$ and $Z$ satisfying also condition C2 and hence $\mA \models_Y \phi_1$ and $\mA \models_Z \phi_2$. This is done in exactly in the same way as in~\cite{10.1145/3471618}.

\end{proof}

In view of Proposition~\ref{prop: weak fragment indep}, \longversion{which shows the hardness of the conjunction of two independence atoms in disjunction with a first-order formula,} this seems to be tight. 

\bibliographystyle{named}

\bibliography{biblio_short}

\end{document}